\documentclass[prd,notitlepage,longbibliography,nofootinbib,superscriptaddress,twocolumn,preprintnumbers]{revtex4-2}
\usepackage[utf8]{inputenc}

\usepackage{bm}
\usepackage{comment} 
\usepackage[colorlinks=true,urlcolor=blue,anchorcolor=black,citecolor=blue,linkcolor=red,filecolor=black,menucolor=black,pagecolor=black,linktocpage=true,pdfproducer=medialab,pdfa=true]{hyperref}
\usepackage{graphicx}
\usepackage{amsmath,latexsym,amssymb,mathrsfs,ascmac,mathtools}
\usepackage{multirow}
\usepackage{braket}

\begin{document}
\preprint{KOBE-COSMO-23-03}
\title{Entropy Bound and a Geometrically Nonsingular Universe}

\author{Takamasa Kanai}
\email{m19013d@math.nagoya-u.ac.jp}
\affiliation{Department of Mathematics, Nagoya University, Nagoya 464-8602, Japan}

\author{Kimihiro Nomura}
\email{knomura@stu.kobe-u.ac.jp}
\affiliation{Department of Physics, Kobe University, Kobe 657-8501, Japan}
 
\author{Daisuke Yoshida}
\email{dyoshida@math.nagoya-u.ac.jp}
\affiliation{Department of Mathematics, Nagoya University, Nagoya 464-8602, Japan}

\begin{abstract}
Bousso's entropy bound is a conjecture that the entropy through a null hypersurface emanating from a two-dimensional surface with a nonpositive expansion is bounded by the area of that two-dimensional surface.
We investigate the validity of Bousso's entropy bound in the spatially flat, homogeneous, and isotropic universe with an adiabatic entropy current. 
We find that the bound is satisfied in the entire spacetime in which a cutoff time is introduced based on the entropy density and the energy density.
Compared to the previously used prescription which puts a cutoff near the curvature singularity, our criterion for introducing the cutoff is applicable even to a nonsingular universe. Our analysis provides an interpretation of the incompleteness implied by the recently proposed singularity theorem based on the entropy bounds. 
\end{abstract}
\maketitle

\section{Introduction}

Singularity theorems \cite{Penrose:1964wq, Hawking:1967ju, Hawking:1970zqf} (see also Refs.~\cite{Hawking:1973uf, Wald:1984rg, o1983semi}) are the most important theorems in general relativity. 
They state that the presence of spacetime singularity is inevitable under assumptions about geodesic convergence (energy condition) and the global structure of spacetime. 
The singularity theorem by Penrose \cite{Penrose:1964wq} is formulated based on a property of null geodesics and hence the timelike convergence condition (strong energy condition) is not required, while the null convergence condition (null energy condition) is assumed.
Hence, the theorem by Penrose is useful to discuss the universe before an inflationary stage where the strong energy condition is violated. Since Penrose's theorem aims to formulate the black hole singularity in asymptotically flat spacetime, the topology of the Cauchy surface is assumed to be noncompact.
Thus, the universe with a compact Cauchy surface can avoid Penrose's theorem.
Properties of the nonsingular universe that evades the assumption of Penrose's theorem are investigated in Ref.~\cite{Borde:1996df} for black holes and in Ref.~\cite{Nomura:2022vcj} for cosmology.

One of the generalizations to relax the assumption of the noncompactness of the Cauchy surface is suggested by Tipler \cite{Tipler:1978zz}, where Penrose's theorem is found to hold when the universal covering of the Cauchy surface is not topologically $S^{3}$. 
Recently, a new kind of singularity theorem is proposed in Ref.~\cite{Bousso:2022cun}.
There the singularity theorem is formulated based on the entropy bounds and the theorem is applicable to any topology of the Cauchy surface.
As a related work on the singularity theorem based on the entropic arguments, see also Ref.~\cite{Wall:2010jtc} for the singularity theorem based on the generalized second law and Ref.~\cite{Bousso:2022tdb} for that based on the quantum Bousso bound. 

The entropy bounds claim that there is an upper bound on the amount of entropy contained in a finite circumstance area.
The first entropy bound, the Bekenstein bound \cite{Bekenstein:1980jp}, is proposed based on the validity of the generalized second law of black hole thermodynamics \cite{Bekenstein:1972tm, Bekenstein:1973ur, Bekenstein:1974ax}. It is proven in free field theory when the gravitational backreaction is negligible \cite{Casini:2008cr}.
A naive generalization of the Bekenstein bound to general spacetimes is called spatial entropy bound.
However, this bound can be easily violated.
Bousso \cite{Bousso:1999xy} proposes that the entropy should be evaluated on null hypersurfaces called light sheets, and this entropy bound is called Bousso bound.
See Refs.~\cite{Wald:1999vt, Bousso:2002ju} for a review of these entropy bounds. 
In the original paper~\cite{Bousso:1999xy}, the Bousso bound is confirmed in the expanding universe with an initial curvature singularity, introducing a cutoff time near the singularity.
See also Refs.~\cite{Fischler:1998st,Kaloper:1999tt} for discussion on the entropy bound in the universe including accelerating one. Proofs of the classical and quantum Bousso bounds with some assumptions are provided in Refs.~\cite{Flanagan:1999jp, Bousso:2014sda}.

The singularity theorem based on the entropy bounds \cite{Bousso:2022cun} basically states that, assuming the global hyperbolicity, the null energy condition, and the validity of the Bousso bound, the violation of the spatial entropy bound leads to the incompleteness of the spacetime.
The assumptions other than the validity of the Bousso bound are satisfied for any spatially flat homogeneous and isotopic universe, that is flat Friedmann--Lema\^{i}tre--Robertson--Walker (FLRW) universe, following the null energy condition.
Thus, for such a universe, the singularity theorem predicts the presence of the geometrical singularity (geodesic incompleteness) of spacetime or the violation of the Bousso bound.
An important note here is that we can construct a geometrically nonsingular (geodesically complete) flat FLRW universe consistent with the null energy condition.
It is possible by considering the maximal extension beyond a coordinate singularity of the inflationary universe \cite{Yoshida:2018ndv, Nomura:2021lzz, Nishii:2021ylb, Nomura:2022vcj}.
The conditions for the past boundary of the inflationary universe predicted by the Borde--Guth--Vilenkin theorem \cite{Borde:2001nh} to be locally extendible are clarified in Refs.~\cite{Yoshida:2018ndv, Nomura:2021lzz, Nishii:2021ylb}.
(A possible inextendibility due to globally nontrivial topology is investigated in Ref.~\cite{Numasawa:2019juw}.)
In the case of the geometrically nonsingular universe, the consequence of the singularity theorem should be incompleteness of the spacetime region satisfying the Bousso bound, not a geometrical singularity, by construction. 
More precisely, the singularity theorem should be applied after introducing some ``cutoff'' in spacetime so that the Bousso bound is satisfied in the entire resultant region, and the theorem implies incompleteness of that region.
In addition, from the consideration of the geometrically nonsingular universe, the ``cutoff'' should be independent of the presence or absence of the geometrical singularity.

In light of the above argument, the main purpose of this paper is to clarify the implication of the singularity theorem based on the entropy bounds \cite{Bousso:2022cun} for the geometrically nonsingular (geodesically complete) universe.
We will propose a criterion for introducing a cutoff time in spacetime based on the local entropy density, not relying on the presence of a geometrical singularity. 
And then, we will check that the Bousso bound is always satisfied in the entire region left after introducing the cutoff.
Once the cutoff is introduced so that the Bousso bound is entirely satisfied, we can apply the singularity theorem even to the geometrically nonsingular universe.
Then, the origin of the incompleteness predicted by the singularity theorem should be understood as the presence of the cutoff introduced due to a large entropy density, not a geometrical singularity.  

This paper is organized as follows.
In the next section, we review the entropy bounds and the singularity theorem based on the entropy bounds formulated in Ref.~\cite{Bousso:2022cun}. 
In Section \ref{sec:3}, we evaluate the entropy-to-area ratio, which is the quantity relevant to the Bousso bound, of adiabatic fluid in the flat FLRW universe.
There, we propose a criterion for a cutoff time that should be introduced so that the Bousso bound is satisfied.
In Section \ref{sec:4}, we explicitly check the Bousso bound is satisfied after introducing the cutoff in the universe with a fluid with a constant equation-of-state parameter. 
Then, in Section \ref{sec:5}, we also check the validity of the Bousso bound in a geometrically nonsingular universe consistent with the null energy condition.
The final section is devoted to the summary and discussion.

Throughout this paper, we use the unit $c = k_{B} = \hbar = 1$, where $c, k_{B}$, and $\hbar$ are the speed of light, the Boltzmann constant, and the reduced Planck constant, respectively.
We assume that spacetime is 4-dimension though the generalization to arbitrary dimension is straightforward.
We use the Newton constant $G$ and the reduced Planck mass $M_{\text{pl}} = (8 \pi G)^{-1/2}$ interchangeably.

\section{Review of Entropy Bounds}

In this section, we review several entropy bounds; the Bekenstein bound \cite{Bekenstein:1980jp}, the spatial entropy bound, and the Bousso bound \cite{Bousso:1999xy}.
Then, we also review the singularity theorem based on these entropy bounds \cite{Bousso:2022cun}.

\subsection{The Bekenstein bound and the spatial entropy bound}

Motivated by an argument on the validity of the generalized second law of black hole thermodynamics \cite{Bekenstein:1972tm,Bekenstein:1973ur,Bekenstein:1974ax}, in Ref.~\cite{Bekenstein:1980jp}, Bekenstein proposed that the amount of entropy $S$ of matter that has the total energy $E$ and is enclosed by a sphere with a radius $R$ has the upper bound given by
\begin{align}
 S \leq 2 \pi E R.\label{Bekensteinbound}
\end{align} 
This inequality does not include the gravitational constant $G$ and hence it is regarded as a property of matter itself, without gravity.  Actually, this inequality is proven for the free field theory when the gravitational backreaction is negligible \cite{Casini:2008cr}.
There, $S$ is defined as the difference of the entanglement entropy by tracing out the outside of the ball from that in the vacuum state. 

In curved spacetime, the notion of the energy $E$ and the radius $R$ of a sphere loses the exact meaning.
To rewrite the Bekenstein bound in a well-defined manner in curved spacetime, let us assume that the size of the system $R$ is larger than the Schwarzschild radius $2 G E$, that is, $E \leq R/(2G)$. 
Then Bekenstein bound can be rephrased as 
\begin{align}
  S \leq \frac{A}{4G},
\end{align}
where $A$ is the area of 2-sphere, $A = 4 \pi R^2$.
This form of inequality is well-defined in curved spacetime and it results in the spatial entropy bound.
In this way, we can consider the following spatial entropy bound as a natural generalization of the Bekenstein bound: 
Let $\Sigma_{\text{in}}$ be a closed, 3-dimensional spacelike hypersurface with a nonvanishing boundary $\sigma \coloneqq  \partial \Sigma_{\text{in}} \neq \varnothing$. Then the spatial entropy bound states that
\begin{align}
 S(\Sigma_{\text{in}}) \leq \frac{A(\sigma)}{4 G}.
 \label{eqseb}
\end{align}  

Throughout this paper, we focus on the entropy of thermal fluid. 
Thus, we assume that there is an entropy current $s^{\mu} \partial_{\mu}$ and the total entropy within a 3-volume $V$ can be defined as the volume integral of the entropy current,
\begin{align}
 S(V) &\coloneqq  \int_{V} s_{3},
\end{align}
where $s_{3}$ is the 3-form defined through the Hodge dual of the entropy current by
\begin{align}
  s_{3} &\coloneqq  - * (s_{\mu} dx^{\mu}).
\end{align}

Once the spacetime geometry and the entropy current are given, the spatial entropy bound \eqref{eqseb} can be tested. 
We call the spatial volume $\Sigma_{\text{in}}$ that violates the spatial entropy bound as {\it hyperentropic hypersurface} as defined in Ref.~\cite{Bousso:2022cun}.

\subsection{Light sheets and the Bousso bound}

The Bousso bound \cite{Bousso:1999xy} is proposed as a possible generalization of the spatial entropy bound.
There, the spatial volume $\Sigma_{\text{in}}$ is replaced with a null hypersurface $L$ called {\it light sheet} defined below.
The Bousso bound states that the entropy through a light sheet $L$ should be bounded by the area of its boundary $\sigma$:
\begin{align}
 S(L) \leq \frac{A(\sigma)}{4 G}.
\end{align}

For $L$ to be a light sheet of $\sigma = \partial \Sigma_{\text{in}} \neq \varnothing$, the following properties are required:
\begin{itemize}
 \item $L$ is a null hypersurface generated by a null geodesic congruence orthogonal to $\sigma$.
\item The congruence of the null generators of $L$ has a nonpositive expansion $\theta \leq 0$ everywhere on $L$. 
\item Each generator of $L$ has an endpoint on $\sigma$.   
\end{itemize}
Here the expansion $\theta$ is defined by $\theta = \nabla_{\mu} k^{\mu}$, letting $k^{\mu}\partial_{\mu}$ be the tangent of null geodesics pointing away from the surface $\sigma$.

There are variations of the definition of the light sheet regarding the other endpoint.
The first version is proposed in the original paper by Bousso~\cite{Bousso:1999xy}.
There, each generator of $L$ is assumed to have an endpoint at the conjugate point ($\theta \rightarrow - \infty$). In this definition, a light sheet can be self-intersecting in general. 
Another version is discussed in Refs.~\cite{Tavakol:1999as, Flanagan:1999jp}. 
There, the generator of $L$ is assumed to have an endpoint at
either the conjugate point or the point where it meets another generator.
We would like to note that, in the proof of the singularity theorem \cite{Bousso:2022cun}, the Bousso bound is evaluated for the light sheet with the latter definition.
Therefore, we will use the latter definition here.
We note that we will focus only on spherically symmetric light sheets in the homogeneous and isotropic universe.
For such light sheets, indeed there is no difference between these definitions.

\subsection{Singularity theorem based on entropy bounds}

The singularity theorem based on the entropy bounds formulated by Ref.~\cite{Bousso:2022cun} can be summarized, for our purpose, as follows:
{\it
Let $({\cal M}, g)$ be a spacetime manifold with a conserved entropy current.  Suppose that the Bousso bound is satisfied for any light sheet.
In addition, let the spacetime satisfy the following properties:
\begin{enumerate}
 \item It is globally hyperbolic. 
 \item It satisfies the null convergence condition.
 \item There is a closed subset $\Sigma_{\mathrm{in}}$ of a Cauchy surface $\Sigma$, that satisfies following three properties:\\
- $\Sigma_{\mathrm{in}}$ has a compact, nonvanishing boundary $\sigma \coloneqq  \partial \Sigma_{\mathrm{in}} \neq \varnothing$ and a nonvanishing interior $\mathrm{int }(\Sigma_{\mathrm{in}}) \neq \varnothing$.\\
- $\Sigma_{\mathrm{in}}$ is a hyperentropic hypersurface: $S(\Sigma_{\mathrm{in}}) > A(\sigma)/4G$.\\
- The future (past) directed inward null geodesic congruence orthogonal to $\sigma$ has a negative expansion on $\sigma$. 
\end{enumerate}
Then the spacetime is future (past) incomplete.
}

Here an inward null geodesic means that the geodesic is toward the direction of $\Sigma_{\text{in}}$.
We note that the spacetime $(\mathcal{M}, g)$ is not assumed to be inextendible here.
Thus, $(\mathcal{M}, g)$ could be a globally hyperbolic subregion of a spacetime. 
In the original paper \cite{Bousso:2022cun}, the definition of entropy is not specified.
The theorem is applicable if the entropy $S$ satisfies the following property: $S(V_{1}) = S(V_{2})$ if $D(V_{1}) = D(V_{2})$, where $D(V_{1})$ represents the domain of dependence of $V_{1}$, and so on. In our case, where the entropy is defined through the entropy current of fluid, this requirement means that the entropy current is conserved.

Contrary to the singularity theorem by Penrose, noncompactness of the Cauchy surface is not required.
Thus, it is applicable to a wider class of geometry.
Instead, we need to assume the presence of nonzero conserved entropy current and the validity of the Bousso bound.
For given spacetime and an entropy current, basically, we are not sure whether the Bousso bound is valid for any light sheet.
Thus, when the above assumptions 1--3 are satisfied, it might mean the violation of the Bousso bound. 

Actually, in examples where the Bousso bound was tested so far (e.g.~the FLRW universe in Ref.~\cite{Bousso:1999xy}), the Bousso bound is confirmed after introducing a cutoff near geometrical curvature singularity.
In this case, the singularity theorem is applicable by regarding $\mathcal{M}$ as spacetime after introducing the cutoff.
Then, the incompleteness predicted by the theorem indicates the presence of the cutoff originated by the violation of the Bousso bound, not the geometrical inextendibility of the spacetime.

Here we should note that, as we will see later, in fact, there exist examples of the geometrically nonsingular flat FLRW universe that satisfy the above assumptions 1--3. 
Since they are geometrically nonsingular, the Bousso bound must be violated as is the case of a universe with curvature singularity.
In the case of the singular universe, the Bousso bound is confirmed after introducing the cutoff time near the singularity where the curvature becomes unity in the Planck unit.
However, contrary to the geometrically singular universe, we cannot introduce the cutoff following the rule based on the curvature because the curvature remains small in the nonsingular universe.
The main purpose of this paper is to clarify a criterion for introducing a cutoff so that the Bousso bound is valid not relying on the curvature, which should be applicable to both singular and nonsingular universes.
Such a criterion will clarify the origin of incompleteness predicted by the singularity theorem based on the entropy bounds. 

\section{The Bousso Bound in the Universe: general analysis}
\label{sec:3}

\subsection{Setup}
Let us focus on an expanding flat FLRW universe where the metric is locally expressed as
\begin{align}
 g_{\mu\nu} dx^{\mu} dx^{\nu} 
 &= - dt^2 + a(t)^2 (d r^2 + r^2 (d \theta^2 + \sin^2 \theta d \phi^2))
 \notag\\
& = a(\eta)^2  \left( - d \eta^2  + d r^2 + r^2 (d \theta^2 + \sin^2 \theta d \phi^2)\right),
\end{align}
where $\eta$ is the conformal time defined through $dt = a d \eta$.
Let us assume the universe is filled with ideal fluid with the energy density $\rho$ and the pressure $p$.
Now Friedmann equations are written as
\begin{align}
 H^2 = \frac{8 \pi G}{3} \rho, \qquad 
 \partial_{t} H = - 4 \pi G (\rho + p),
 \label{friedmann}
\end{align}
where $H = \partial_t a / a$.

We also assume that the fluid has thermodynamic entropy that is characterized by an entropy current
\begin{align}
 s^{\mu} \partial_{\mu} = s(t) \partial_{t}.
\end{align}
The total entropy within a 3-volume $V$ is defined as
\begin{align}
 S(V) &= \int_{V} s_{3},
\end{align}
where $s_{3}$ is calculated as
\begin{align}
  s_{3} & = s(t) a(t)^3 r^2 \sin \theta \,  dr \wedge d\theta \wedge d\phi.
\end{align}
Throughout this paper, we assume that the expansion of the universe is adiabatic; that is, we assume
\begin{align}
 s(t) = \frac{\bar{s}}{a(t)^3}
\end{align}
with a constant $\bar{s}$. With this assumption, the entropy current is conserved $\nabla_{\mu} s^{\mu} = 0$ and the singularity theorem based on the entropy bounds is applicable. When the 3-dimensional volume $V$ is parameterized by $r,\theta,\phi$ as $t = t(r), r \in [r_{1}, r_{2}], \theta \in [0, \pi], \phi \in [0, 2 \pi)$, the entropy can be evaluated as 
\begin{align}
 S(V) 
 &= 4 \pi \int_{r_{1}}^{r_{2}} s(t(r)) a(t(r))^3 r^2 dr \notag\\
 &= \bar{s} \cdot \frac{4 \pi}{3} (r_{2}^3- r_{1}^3).
 \label{SV}
\end{align}
Here we used the fact that $s(t)a(t)^3$ is equal to the constant $\bar{s}$.

\subsection{Entropy-to-area ratio for spatial hypersurfaces}

Let us focus on a 2-dimensional sphere $\sigma$ defined by $\eta = \eta_{0}$ and $r = r_{0}$. Let us call the inside ($0< r < r_{0}$) and the outside ($r_{0} < r$) of $\sigma$ on the Cauchy surface $\Sigma$ as $\Sigma_{\text{in}}$ and $\Sigma_{\text{out}}$, respectively. 
We call the direction towards $\Sigma_{\text{in}}$ as ingoing, and that towards $\Sigma_{\text{out}}$ as outgoing. Now we can check the spatial entropy bound for $\Sigma_{\text{in}}$ and $\Sigma_{\text{out}}$. The total entropy within the hypersurface $\Sigma_{\text{in}}$ can be estimated as
\begin{align}
 S(\Sigma_{\text{in}})  = \bar{s} \cdot
 \frac{4}{3} \pi r_{0}^3,
\end{align}
whereas the physical area of the surface $\sigma$ is written as
\begin{align}
 A(\sigma) = a(\eta_{0})^2 \cdot 4 \pi r_{0}^2.\label{Asigma}
\end{align}
Let us introduce the entropy-to-area ratio for the spatial entropy bound $\mathscr{R}_{\text{sp}}$ by
\begin{align}
 \mathscr{R}_{\text{sp}}(r_{0}, \eta_{0}) \coloneqq   \frac{S(\Sigma_{\text{in}})}{A(\sigma)/4G}.\label{Rsp}
\end{align}
We obtain 
\begin{align}
\mathscr{R}_{\text{sp}}(r_{0}, \eta_{0})  = \frac{1}{6\pi} \frac{s(\eta_{0})}{M_{\text{pl}}^3} 
\frac{a(\eta_{0}) r_{0}}{M_{\text{pl}}^{-1}},
\label{Rspex}
\end{align}
and since it is proportional to $r_{0}$,  the spatial entropy bound is violated for the inside of a sufficiently large sphere.

Similarly, since the entropy on $\Sigma_{\text{out}}$ diverges, the Bekenstein bound for the outside of any sphere is violated:
\begin{align}
 \frac{S(\Sigma_{\text{out}})}{A(\sigma)/4G} = \infty.
\end{align}
Thus, $\Sigma_{\text{in}}$ and $\Sigma_{\text{out}}$ can be hyperentropic hypersurfaces.

\subsection{Light sheets}

Suppose that the null convergence condition is satisfied; $R_{\mu\nu} k^\mu k^\nu \geq 0$ for any null vector $k^\mu$, where $R_{\mu\nu}$ is the Ricci tensor.
A null surface generated by null geodesics which are orthogonal to $\sigma$ and initially converging on $\sigma$ becomes a light sheet from $\sigma$ because the expansion on it is always negative. 
Let us calculate the initial expansion on $\sigma$ for all the possible directions, future-outgoing, future-ingoing, past-outgoing, and past-ingoing, which are indicated by the suffixes ``f,o'', ``f,i'', ``p,o'', and ``p,i'', respectively. 
The tangent of the affine parametrized null geodesics orthogonal to $\sigma$ can be expressed as
\begin{align}
 k^{\mu}_{\text{f},\text{o}} \partial_{\mu} &=  \frac{1}{a(t)} \partial_{t} + \frac{1}{a(t)^2} \partial_{r}, \\
 k^{\mu}_{\text{f},\text{i}} \partial_{\mu} &=   \frac{1}{a(t)} \partial_{t} - \frac{1}{a(t)^2} \partial_{r}, \\
 k^{\mu}_{\text{p},\text{o}} \partial_{\mu} &=  -  \frac{1}{a(t)} \partial_{t} + \frac{1}{a(t)^2} \partial_{r}, \\
 k^{\mu}_{\text{p},\text{i}} \partial_{\mu} &=  -  \frac{1}{a(t)} \partial_{t} - \frac{1}{a(t)^2} \partial_{r}.
\end{align}
As explained above, the terminology of outgoing/ingoing is based on the $r$ coordinate.
Note that $k_{\text{f,o}}^{\mu} = - k_{\text{p,i}}^{\mu}$ and $k_{\text{f,i}}^{\mu} = - k_{\text{p,o}}^{\mu}$. The expansion of each congruence of null geodesics is evaluated as \cite{Nomura:2022vcj}
\begin{align}
 \theta_{\text{f},\text{o}} &= \frac{2}{a} \left(  H + \frac{1}{a r}\right), \\
\theta_{\text{f},\text{i}} &= \frac{2}{a} \left( H -  \frac{1}{a r}\right), \\
\theta_{\text{p},\text{o}} &= \frac{2}{a} \left( - H + \frac{1}{a r}\right), \\
\theta_{\text{p},\text{i}} &= \frac{2}{a} \left( -  H -  \frac{1}{a r}\right).
\end{align}
Since we focus on the expanding universe $H>0$, $\theta_{\text{f},\text{o}}$ is always positive and $\theta_{\text{p},\text{i}}$ is always negative. $\theta_{\text{f},\text{i}}$ and $\theta_{\text{p},\text{o}}$ change their sign at the cosmological apparent horizon
\begin{align}
r_{\text{AH}}(\eta) \coloneqq  \frac{1}{a(\eta)H(\eta)}.
\end{align}
For the inside of the apparent horizon ($r_{0} \leq r_{\text{AH}}(\eta_{0})$), the ingoing directions (future-ingoing and past-ingoing) are converging as in flat spacetime because the effect of the expansion of the universe is week enough.
On the other hand, for the outside of the apparent horizon ($r_{\text{AH}}(\eta_{0}) \leq r_{0}$), the future directions are expanding due to the expansion of the universe; in other words, both the past directions (past-ingoing and past-outgoing) are converging.
Hence, there are future-ingoing light sheet $L_{\text{f},\text{i}}$ and past-ingoing light sheet $L_{\text{p},\text{i}}$ inside the apparent horizon $(r_{0} \leq r_{\text{AH}}(\eta_{0}))$.
On the other hand, there are past-ingoing light sheet $L_{\text{f},\text{i}}$ and past-outgoing light sheet $L_{\text{p},\text{i}}$ outside the apparent horizon $(r_{\text{AH}}(\eta_{0}) \leq r_{0})$.

The singularity theorem is applicable when the hyper entropic hypersurface is located in a trapped direction of $\sigma$. For a sufficiently large sphere that is larger than the apparent horizon, there are both ingoing and outgoing light sheets, and hence the hyperentropic regions $\Sigma_{\text{in}}$ and $\Sigma_{\text{out}}$ satisfy the conditions in the singularity theorem. On the one hand, if one considers a sphere smaller than the apparent horizon, $\Sigma_{\text{out}}$ is actually hyperentropic but there is no outgoing light sheet.
Such a hypersurface does not satisfy the assumption in the singularity theorem.

\subsection{Entropy-to-area ratio for light sheets}

Let us evaluate the entropy-to-area ratio in a general setting where the conformal time is defined in $\eta \in (\eta_{\text{i}}, \eta_{\text{f}})$.
We note that $\eta_{\text{i}}$ and $\eta_{\text{f}}$ could be infinite.

In case where $\eta_{\text{i}}$ is finite, we can define the particle horizon by
\begin{align}
 r_{\text{PH}}(\eta) = \eta - \eta_{\text{i}}.
\end{align}
For $r_{\text{PH}} < r_{0}$, a past-ingoing light sheet $L_{\text{p},\text{i}}$ hits the $\eta = \eta_{\text{i}}$ hypersurface.
  
Similarly, when $\eta_{\text{f}}$ is finite, we can define the event horizon by
\begin{align}
 r_{\text{EH}}(\eta) = - \eta + \eta_{\text{f}}.
\end{align}
For $r_{\text{EH}} < r_{0}$, a future-ingoing light sheet $L_{\text{f},\text{i}}$ hits the $\eta = \eta_{\text{f}}$ hypersurface.

When $\eta_{\text{i}} = - \infty$, there is no particle horizon. In this case we regard $r_{\text{PH}} = \infty$.
Similarly, when $\eta_{\text{f}} = + \infty$, there is no event horizon and we regard $r_{\text{EH}} = \infty$.

When the value of the $r$ coordinate of a light sheet runs $r_{1} \leq r \leq r_{2}$, from Eqs.~\eqref{SV} and \eqref{Asigma} the entropy-to-area ratio can be evaluated as
\begin{align}
 \frac{S(L)}{A(\sigma)/4G} =  \frac{1}{6\pi}  \frac{s(\eta_{0})}{M_{\text{pl}}^3}  \frac{a(\eta_{0}) r_{0}}{M_{\text{pl}}^{-1}} \frac{r_{2}^3 - r_{1}^3}{r_{0}^3} .
\end{align}

\subsubsection{Past-ingoing light sheets $L_{\mathrm{p,i}}$}

The past-ingoing light sheet $L_{\text{p},\text{i}}$ exists for any value of $r_{0}$.
Let us calculate the entropy-to-area ratio for a past-ingoing light sheet 
\begin{align}
 \mathscr{R}_{\text{p,i}}(r_{0}, \eta_{0}) \coloneqq  \frac{S(L_{\text{p,i}})}{A(\sigma)/4G}.
\end{align}

For $r_{0} \leq r_{\text{PH}}$, $L_{\text{p,i}}$ does not intersect with the initial $\eta = \eta_{\text{i}}$ hypersurface and hence the coordinate value $r$ of the light sheet runs $r \in [0, r_{0}]$.
Thus, the entropy-to-area ratio of the past-ingoing light sheet $\mathscr{R}_{\text{p,i}}$ is evaluated as
\begin{align}
 \mathscr{R}_{\text{p,i}}(r_{0}, \eta_{0}) = \frac{1}{6\pi} \frac{s(\eta_{0})}{M_{\text{pl}}^3} \frac{a(\eta_{0}) r_{0}}{M_{\text{pl}}^{-1}},
\end{align}
for $r_{0} \leq r_{\text{PH}}(\eta_{0})$.
The maximum value is obtained when $r_{0} = r_{\text{PH}}(\eta_{0})$.
Note that if there is no particle horizon $r_{\text{PH}} = \infty$, the entropy-to-area ratio of a past-ingoing light sheet is unbounded. Thus, in order to satisfy the Bousso bound, $\eta_{\text{i}}$ must be finite.

For $r_{\text{PH}}(\eta_{0}) < r_{0}$, $L_{\text{p,i}}$ touches to the initial time slice $\eta = \eta_{\text{i}}$ and hence the coordinate value $r$ runs  $r_{0} - r_{\text{PH}}(\eta_{0}) \leq r \leq r_{0}$.
Thus, the entropy-to-area ratio can be evaluated as
\begin{align}
\mathscr{R}_{\text{p,i}}(r_{0}, \eta_{0}) 
&= \frac{1}{6\pi}\frac{s(\eta_{0})}{M_{\text{pl}}^3} \frac{a(\eta_{0}) r_{0}}{M_{\text{pl}}^{-1}}\frac{ r_{0}^3 - (r_{0} - r_{\text{PH}})^3}{r_{0}^3}  \notag\\
&= 
\frac{1}{6\pi}\frac{s(\eta_{0})}{M_{\text{pl}}^3} \frac{a(\eta_{0}) r_{\text{PH}}}{M_{\text{pl}}^{-1}}
 \left( 3  - 3  \frac{r_{\text{PH}}}{r_{0}} + \frac{r_{\text{PH}}^2}{r_{0}^2} \right)
\end{align}
for $r_{\text{PH}}(\eta_{0}) < r_{0}$.
Noting that $0 <r_{\text{PH}}/r_{0} \leq 1$, it takes the maximum value when $r_{0} \rightarrow \infty $.

The maximum value of the entropy-to-area ratio of past-ingoing light sheets from the $\eta = \eta_0$ slice
\begin{align}
  \mathscr{R}_{\text{p,i}}^{\max}(\eta_{0}) \coloneqq  \max_{r_{0} \in [0, \infty)} \mathscr{R}_{\text{p,i}}(r_{0}, \eta_{0})
\end{align}
can be evaluated as
\begin{align}
\mathscr{R}_{\text{p,i}}^{\max}(\eta_{0}) &= \mathscr{R}_{\text{p,i}}( \infty, \eta_{0}) \notag\\
& = 
\frac{1}{6\pi}\frac{s(\eta_{0})}{M_{\text{pl}}^3} \frac{a(\eta_{0}) r_{\text{AH}}(\eta_{0})}{M_{\text{pl}}^{-1}}
 \cdot 3 \frac{r_{\text{PH}}(\eta_{0})}{r_{\text{AH}}(\eta_{0})}.
\label{SoverAmaxpi}
\end{align} 
By the definition, if $\mathscr{R}_{\text{p,i}}^{\text{max}}(\eta_{0}) \leq 1$, all the past-ingoing light sheets starting from the $\eta = \eta_{0}$ slice satisfy the Bousso bound, $\mathscr{R}_{\text{p,i}}(r_{0}, \eta_{0}) \leq 1$.

\subsubsection{Future-ingoing light sheets $L_{\mathrm{f,i}}$}

The future-ingoing light sheet exists if $r_{0} \leq r_{\text{AH}}(\eta_{0})$. 
Let us calculate the entropy-to-area ratio for a future-ingoing light sheet  defined by
\begin{align}
 \mathscr{R}_{\text{f,i}}(r_{0}, \eta_{0}) \coloneqq  \frac{S(L_{\text{f,i}})}{A(\sigma)/4G}.
\end{align}
Below, we consider following two cases separately: $r_{\text{AH}} \leq r_{\text{EH}}$ and $r_{\text{EH}} < r_{\text{AH}}$.

In the case with $r_{\text{AH}} \leq r_{\text{EH}}$, for any value of $r_{0} \leq r_{\text{AH}}(\leq r_{\text{EH}})$, 
the entropy-to-area ratio for a future-ingoing light sheet is given by
\begin{align}
\mathscr{R}_{\text{f,i}}(r_{0}, \eta_{0})  = \frac{1}{6\pi} \frac{s(\eta_{0})}{M_{\text{pl}}^3} \frac{a(\eta_{0}) r_{0}}{M_{\text{pl}}^{-1}}.
\end{align}
It takes the maximum value when $r_{0} = r_{\text{AH}}$.

In the case with $r_{\text{EH}} < r_{\text{AH}}$, there are two kinds of future-ingoing light sheets depending on the radius. 
One is with $r_{0} \leq r_{\text{EH}}(< r_{\text{AH}})$, where the entropy-to-area ratio can be evaluated as the above case:
\begin{align}
\mathscr{R}_{\text{f,i}}(r_{0}, \eta_{0}) = 
\frac{1}{6\pi} \frac{s(\eta_{0})}{M_{\text{pl}}^3} \frac{a(\eta_{0}) r_{0}}{M_{\text{pl}}^{-1}}.
\end{align}
It takes the maximum value when $r_{0} = r_{\text{EH}}.$ 
The other is the light sheets with $r_{\text{EH}} < r_{0} \leq r_{\text{AH}}$, where each light sheet touches the time slice $\eta = \eta_{\text{f}}$ and hence the $r$ coordinate runs $ r_{0} - (\eta_{\text{f}} - \eta_{0}) \leq r \leq r_{0}$. 
Then, the entropy-to-area ratio can be evaluated as
\begin{align}
 \mathscr{R}_{\text{f,i}}(r_{0}, \eta_{0}) &= \frac{1}{6 \pi} \frac{s(\eta_{0})}{M_{\text{pl}}^3} \frac{a(\eta_{0}) r_{0}}{M_{\text{pl}}^{-1}} \frac{  r_{0}^3 - (r_{0} - ( \eta_{\text{f}} - \eta_{0}) )^3  }{r_{0}^3} \notag\\
&= \frac{1}{6 \pi} \frac{s(\eta_{0})}{M_{\text{pl}}^3} \frac{a(\eta_{0}) r_{\text{EH}}}{M_{\text{pl}}^{-1}}  \left( 3 - 3 \frac{r_{\text{EH}}}{r_{0}}  + \frac{r_{\text{EH}}^2}{r_{0}^2} \right).
\end{align}
Noting that
\begin{align}
 0 \leq  \frac{r_{\text{EH}}}{r_{\text{AH}}} \leq \frac{r_{\text{EH}}}{r_{0}}  \leq 1,
\end{align}
the maximum value is obtained when $r_{0} = r_{\text{AH}}$.
The entropy-to-area ratio of the light sheet from $r_{0} = r_{\text{AH}}$ surface is greater than that from $r_{0} = r_{\text{EH}}$.
Thus the maximum value of the entropy-to-area ratio is given at $r_{0} = r_{\text{AH}}$ even in the $r_{\text{EH}} < r_{\text{AH}}$ case. 

To summarize, letting $\mathscr{R}^{\text{max}}_{\text{f,i}}(\eta_{0})$ be the maximum value of the entropy-to-area ratio for future-ingoing light sheets from the time slice $\eta = \eta_{0}$, 
\begin{align}
 \mathscr{R}^{\text{max}}_{\text{f,i}}(\eta_{0}) 
 \coloneqq  \max_{r_{0} \in [0, r_{\text{AH}}(\eta_{0})]} 
 \mathscr{R}_{\text{f,i}}(r_{0}, \eta_{0}),
\end{align}
it is given by the value at the apparent horizon $r_{\text{AH}}$,
\begin{widetext}
\begin{align}
\mathscr{R}_{\text{f,i}}^{\text{max}}(\eta_{0}) &= \mathscr{R}_{\text{f,i}}(r_{\text{AH}}(\eta_{0}), \eta_{0}) \notag\\
& =
\frac{1}{6 \pi} \frac{s(\eta_{0})}{M_{\text{pl}}^3} \frac{a(\eta_{0}) r_{\text{AH}}(\eta_{0})}{M_{\text{pl}}^{-1}} \times
\begin{dcases}
 1, 
 & (r_{\text{AH}} \leq r_{\text{EH}})\\
 \frac{r_{\text{EH}}}{r_{\text{AH}}} \left(3 - 3 \left(\frac{r_{\text{EH}}}{r_{\text{AH}}}\right) + \left( \frac{r_{\text{EH}}}{r_{\text{AH}}} \right)^2 \right),
 & (r_{\text{EH}} < r_{\text{AH}}).
\end{dcases}
\label{SoverAmaxfi}
\end{align}
\end{widetext}

\subsubsection{Past-outgoing light sheets $L_{\mathrm{p,o}}$}

The past-outgoing light sheet exists if $r_{\text{AH}} \leq r_{0}$. 
Let us calculate the entropy-to-area ratio for a past-outgoing light sheet defined by
\begin{align}
 \mathscr{R}_{\text{p,o}}(r_{0}, \eta_{0}) \coloneqq   \frac{S(L_{\text{p,o}})}{A(\sigma)/4G}.
\end{align} 
Since the light sheet touches the initial time slice $\eta = \eta_{\text{i}}$, the coordinate value $r$ runs $r_{0} \leq r \leq r_{0}+ \eta_{0} - \eta_{\text{i}}$.
The entropy-to-area ratio can be evaluated as
\begin{align}
 \mathscr{R}_{\text{p,o}}(r_{0},\eta_{0}) &= \frac{1}{6\pi} \frac{s(\eta_{0})}{M_{\text{pl}}^3} \frac{a(\eta_{0}) r_{0}}{M_{\text{pl}}^{-1}} \frac{(r_{0} + (\eta_{0} - \eta_{\text{i}}))^3 -  r_{0}^3}{r_{0}^3} \notag\\
&= \frac{1}{6\pi} \frac{s(\eta_{0})}{M_{\text{pl}}^3} \frac{a(\eta_{0}) r_{\text{PH}}}{M_{\text{pl}}^{-1}}  \left( 3  + 3  \frac{r_{\text{PH}}}{r_{0}} + \frac{r_{\text{PH}}^2}{r_{0}^2} \right),
\end{align}
for $r_{\text{AH}} \leq r_{0}$.
It takes the maximum value at $r_{0} = r_{\text{AH}}$. 
Thus, by defining the maximum value of the entropy-to-area ratio for the past-outgoing light sheets from the $\eta = \eta_{0}$ slice by
\begin{align}
 \mathscr{R}^{\text{max}}_{\text{p,o}}(\eta_{0}) \coloneqq  \max_{r_{0} \in [r_{\text{AH}}(\eta_{0}), \infty)} \mathscr{R}_{\text{p,o}}(r_{0}, \eta_{0}),
\end{align}
it can be evaluated as
\begin{align}
 \mathscr{R}^{\text{max}}_{\text{p,o}}(\eta_{0})  &=  \mathscr{R}_{\text{p,o}}(r_{\text{AH}}(\eta_{0}), \eta_{0})  \notag\\
& =
\frac{1}{6\pi} \frac{s(\eta_{0})}{M_{\text{pl}}^3} \frac{a(\eta_{0}) r_{\text{AH}}(\eta_{0})}{M_{\text{pl}}^{-1}} 
\notag \\
&\quad \times \frac{r_{\text{PH}}}{r_{\text{AH}} }\left(3 + 3 \left( \frac{r_{\text{PH}}}{r_{\text{AH}}}\right) + \left( \frac{r_{\text{PH}}}{r_{\text{AH}}}\right)^2 \right). \label{SoverAmaxpo}
\end{align}

\subsubsection{Summary of the calculations}

Summarizing the above results, the maximum value of the entropy-to-area ratio for the past-ingoing light sheets, future-ingoing light sheets, and past-outgoing light sheets from the $\eta = \eta_{0}$ surface are given by Eq.~\eqref{SoverAmaxpi}, Eq.~\eqref{SoverAmaxfi}, and Eq.~\eqref{SoverAmaxpo}, respectively. 
In order to satisfy the Bousso bound, all of them should be smaller than $1$. Let us define the maximum value of the entropy-to-area ratio on the $\eta = \eta_{0}$ slice by
\begin{align}
 \mathscr{R}^{\text{max}}(\eta_{0}) \coloneqq 
 \max \{ \mathscr{R}^{\text{max}}_{\text{p,i}}(\eta_{0}), \mathscr{R}^{\text{max}}_{\text{f,i}}(\eta_{0}), \mathscr{R}^{\text{max}}_{\text{p,o}}(\eta_{0}) \}.
\end{align}
The Bousso bound on the $\eta = \eta_{0}$ slice can be rephrased as $\mathscr{R}^{\text{max}}(\eta_{0}) \leq 1$.
Comparing Eq.~\eqref{SoverAmaxpi} with Eq.~\eqref{SoverAmaxpo}, the maximum value for the past-ingoing light sheets is always smaller than that for the past-outgoing light sheets, 
$\mathscr{R}_{\text{p,i}}^{\text{max}}(\eta_{0}) <  \mathscr{R}_{\text{p,o}}^{\text{max}}(\eta_{0})$. 
Thus the maximum value of the entropy-to-area ratio is given by either that of the future-ingoing light sheets or that of the past-outgoing light sheets.

Next, let $\mathscr{R}^{\text{AH}}_{\text{sp}}(\eta_{0})$ be the entropy-to-area ratio with respect to the spatial volume $\Sigma_{\text{in}}$ with $r_{0} = r_{\text{AH}}$.
From the definition \eqref{Rsp}, $\mathscr{R}^{\text{AH}}_{\text{sp}}(\eta_{0})$ reads 
\begin{align}
\mathscr{R}_{\text{sp}}^{\text{AH}}(\eta_{0}) &\coloneqq  \mathscr{R}_{\text{sp}}(r_{\text{AH}}(\eta_{0}), \eta_{0}) \notag\\
& = \frac{1}{6\pi} \frac{s(\eta_{0})}{M_{\text{pl}}^3} \frac{a(\eta_{0}) r_{\text{AH}}(\eta_{0})}{M_{\text{pl}}^{-1}}.
\end{align}
Then, $\mathscr{R}^{\text{max}}(\eta_{0})$ can be represented as
\begin{widetext}
\begin{align}
 \mathscr{R}^{\text{max}}(\eta_{0}) 
&= 
\mathscr{R}^{\text{AH}}_{\text{sp}}(\eta_{0})
\times 
\begin{dcases}
\text{max}\left\{1, \frac{r_{\text{PH}}}{r_{\text{AH}}} \left(3 + 3 \left(\frac{r_{\text{PH}}}{r_{\text{AH}}}\right) + \left(\frac{r_{\text{PH}}}{r_{\text{AH}}}\right)^2 \right) \right\},   
& (r_{\text{AH}} \leq r_{\text{EH}}),\\
\text{max}\left\{\frac{r_{\text{EH}}}{r_{\text{AH}}} \left(3 - 3 \left(\frac{r_{\text{EH}}}{r_{\text{AH}}}\right) + \left(\frac{r_{\text{EH}}}{r_{\text{AH}}}\right)^2 \right), \frac{r_{\text{PH}}}{r_{\text{AH}}} \left(3 + 3 \left(\frac{r_{\text{PH}}}{r_{\text{AH}}}\right) + \left(\frac{r_{\text{PH}}}{r_{\text{AH}}}\right)^2 \right) \right\},  
&  (r_{\text{EH}} < r_{\text{AH}}) .
\end{dcases}
\label{Reta0}
\end{align}
\end{widetext}

\subsection{Our criterion for the cutoff time}
The important observation here is that $\mathscr{R}_{\text{sp}}^{\text{AH}}$ can be expressed by the local quantity of the matter contents, the entropy density $s$ and the energy density $\rho$, 
\begin{align}
 \mathscr{R}_{\text{sp}}^{\text{AH}}(\eta_{0}) = \frac{1}{6\pi} \frac{s(\eta_{0})}{M_{\text{pl}}^3} \frac{M_{\text{pl}}}{H(\eta_{0})}
 = \frac{1}{2 \sqrt{3} \pi} \frac{s(\eta_{0})}{M_{\text{pl}}^3} \sqrt{\frac{M_{\text{pl}}^4}{\rho(\eta_{0})}},
\end{align}
where we used the Friedmann equation in the expanding universe \eqref{friedmann}. 
Also, note that the inequality 
\begin{align}
    \frac{r_{\text{PH}}}{r_{\text{AH}}} \left(3 + 3 \left(\frac{r_{\text{PH}}}{r_{\text{AH}}}\right) + \left(\frac{r_{\text{PH}}}{r_{\text{AH}}}\right)^2 \right) 
    \leq 1
\end{align}
is satisfied for $r_{\text{PH}}/r_{\text{AH}} \leq q_{*}  \simeq  0.260$, where $q_{*}$ is defined as the positive root of $q_{*}(3 + 3 q_{*} + q_{*}^2) = 1$.
Then, we can see that at least on the time slice $\eta = \eta_0$ satisfying $r_{\text{PH}}(\eta_{0})/r_{\text{AH}}(\eta_{0}) \leq q_{*} \simeq  0.260$, i.e., the time slice where the particle horizon is sufficiently smaller than the apparent horizon, the Bousso bound is guaranteed by assuming the local condition $\mathscr{R}_{\text{sp}}^{\text{AH}}(\eta_{0}) \leq 1$ because $\mathscr{R}^{\text{max}}(\eta_{0}) = \mathscr{R}_{\text{sp}}^{\text{AH}}(\eta_{0})$ in this case.
Motivating by this fact, we shall propose the condition
\begin{align}
\mathscr{R}_{\text{sp}}^{\text{AH}}(\eta_{\text{cut}}) =  \frac{1}{2 \sqrt{3} \pi}\frac{ s(\eta_{\text{cut}})/M_{\text{pl}}^3}{\sqrt{\rho(\eta_{\text{cut}})/M_{\text{pl}}^4}} = {\cal O}(1)  \label{ourcriterion}
\end{align}
as our criterion for introducing the cutoff time $\eta_{\text{cut}}$ in spacetime so that the Bousso bound is satisfied in the resultant region.
Depending on the situation, it might be a past cutoff $\eta_{\text{cut}} = \eta_{\text{i}}$ or a future cutoff $\eta_{\text{cut}} = \eta_{\text{f}}$.
We will expect that the description of classical gravity with entropic fluid is valid only when $\mathscr{R}_{\text{sp}}^{\text{AH}}(\eta_{0}) \leq {\cal O}(1)$. 
In the later sections, we will explicitly check whether the Bousso bound is actually satisfied in the entire region left after introducing our cutoff \eqref{ourcriterion}.

We note that the behavior of $\mathscr{R}_{\text{sp}}^{\text{AH}}$ is controlled by the dominant energy condition because $\mathscr{R}_{\text{sp}}^{\text{AH}}(\eta) \propto 1/ (a^3 H)$ and the time derivative can be evaluated as
\begin{align}
\partial_{\eta} \mathscr{R}_{\text{sp}}^{\text{AH}}(\eta) &\propto \partial_{\eta} \left( \frac{1}{a^3 H}\right) \notag\\
&= \frac{- 3 H^2 -  \partial_{t}H}{a^2 H^2} 
\notag \\
&= \frac{ - 4 \pi G (\rho - p)}{a^2 H^2}.
\end{align}
Thus, $\mathscr{R}_{\text{sp}}^{\text{AH}}$ is decreasing in time if $\rho > p$, which follows from the dominant energy condition $\rho > |p|$.
In the case where $\rho > p$ is satisfied in the arbitrary past, we possibly need to introduce the initial cutoff $\eta_{\text{i}}$ based on our criterion.
On the one hand, if $\rho < p$ is satisfied in the arbitrary future, we possibly need to introduce the final cutoff $\eta_{\text{f}}$.

Let us see the relations between our criterion and others used in the previous studies.
In Refs.~\cite{Bousso:1999xy, Kaloper:1999tt}, the cutoff time is introduced because of the curvature singularity.
Thus, the classical description of spacetime is considered to be reliable until the energy density (or curvature of the spacetime) approaches the Planck scale and the cutoff time $\eta_{\text{i}}$ is introduced by $\rho(\eta_{\text{i}})/M_{\text{pl}}^{4} \sim {\cal O}(1)$. 
In addition, the entropy density on the cutoff surface is assumed to be of the order $1$ in the Planck unit, $ s(\eta_{\text{i}})/M_{\text{pl}}^3 \sim {\cal O}(1)$. 
Our criterion includes this previous criterion as a special case where $\rho(\eta_{\text{i}})/ M_{\text{pl}}^4 \sim {\cal O}(1)$. 
One merit of our criterion is that it is applicable even when the energy density remains smaller than the Planck scale but the entropy density diverges. This actually happens if one considers a geometrically nonsingular universe that is consistent with the null energy condition, as we will see later.

\section{Constant Equation of State}
\label{sec:4}
Let us focus on the case where the equation of state of the fluid is given by $p = w \rho$ with a constant $w$. 
We assume $w \neq -1/3$ for now, and the case $w = -1/3$ is investigated later.
The conservation law of the energy-momentum tensor can be solved as
\begin{align}
 \rho &= \bar{\rho} \left(\frac{1}{a} \right)^{3 (1 + w)} = \bar{\rho} \left( \frac{1}{a} \right)^{2 (1+q)/ q},
\end{align}
where $\bar{\rho}$ is the energy density per unit comoving volume and we introduce a constant $q$ by
\begin{align}
 q = \frac{2}{1 + 3 w}.
\end{align}
The correspondence between $w$ and $q$ is summarized in Table \ref{TBL1}.

\begin{table*}[htbp]
\begin{tabular}{l||c|c|c|c|c|c|c|c|c|c|c|c|c|c}
$w$ & $-\infty$ & $\cdots$ & $-1$ & $\cdots$ & $-1/3-0$ & $-1/3+0$ &$\cdots$ & $0$ & $\cdots$ & $1/3$ & $\cdots$ & $1$& $\cdots$ & $+ \infty$ \\
\hline
$q$  & $- 0$ & $\searrow$ & $-1$ & $\searrow$ & $- \infty$& $+ \infty$ &$\searrow$ & $2$ &$\searrow$ &  $1$ &$\searrow$ &  $1/2$ &$\searrow$ &  $+0$ 
\end{tabular}
\caption{The correspondence between $w$ and $q$.}
\label{TBL1}
\end{table*}

The Friedmann equation can be solved as
\begin{align}
 a &=
   \left(\frac{\eta}{\bar{\eta}} \right)^{q},
\end{align}
with defining $\bar{\eta}$ by
\begin{align}
 \bar{\eta} =  q \sqrt{ \frac{3}{8 \pi G \bar{\rho}}}.
\end{align}
Here we fix the origin of the conformal time so that $\eta = \bar{\eta}$ corresponds to $a = 1$. By this definition, the smooth spacetime is defined in $\eta \in (0, \infty)$ for $q >0$ and in $\eta \in (- \infty, 0)$ for $q < 0$, that means the spacetime is conformally isometric to the upper half of the Minkowski space for $q > 0$ and the lower half of the Minkowski space for $q < 0$.

The Hubble parameter can be evaluated as 
\begin{align}
 H = \frac{\partial_{\eta}a(\eta)}{a(\eta)^2} = \frac{q}{\bar{\eta}} \left( \frac{\eta}{\bar{\eta}}\right)^{-(1 +q)}.
\end{align}
The Hubble parameter $H$ diverges at $\eta = 0$ when $q + 1 > 0 $, which corresponds to $ w < -1$ or $ - 1/3 < w$. $H$ diverges at $\eta = - \infty$ when $q + 1 < 0$, which corresponds to $ -1 < w < - 1/3$. 
These cases correspond to the scalar curvature singularity. 
The quantity $\partial_{t}H/a^2$ diverges at $\eta = - \infty$ when $q < -1/2$ with $q \neq -1$, which corresponds to $-5/3 < w < -1, -1 < w < - 1/3$. This case corresponds to (nonscalar) curvature singularity~\cite{Yoshida:2018ndv, Nomura:2021lzz,Nishii:2021ylb}. 
See also Ref.~\cite{Harada:2021yul} for the general analysis of the structure of the FLRW universe with a constant equation of state. 

Since $a H  = \partial_{\eta} \log a = q /\eta$, the apparent horizon can be evaluated as 
\begin{align}
 r_{\text{AH}}(\eta) = \frac{\eta}{q}.
\end{align}
The apparent horizon is timelike when $|q| > 1$ $(-1 < w < 1/3)$ , null when $|q| = 1$ $(w = -1, 1/3 )$, and spacelike when $|q| < 1$ $(w < -1, 1/3 < w)$.  

Since $\mathscr{R}_{\text{sp}}^{\text{AH}}(\eta)$ scales as
\begin{align}
\mathscr{R}_{\text{sp}}^{\text{AH}}(\eta) \propto \frac{1}{a(\eta)^{(2 q - 1)/q}} = \frac{1}{a(\eta)^{3(1 - w)/2}},
\label{RspAHs}
\end{align}
and our criterion requires $\mathscr{R}_{\text{sp}}^{\text{AH}}(\eta) < {\cal O}(1)$ for a spacetime region in which the description of the system is trustable, we need to introduce a past cutoff for $w < 1$ and a future cutoff for $w > 1$. 

From the first Friedmann equation \eqref{friedmann}, the energy density of the fluid $\rho$ must be positive. 
Then, to satisfy the null energy condition, which is one of the assumptions of the singularity theorem, we must have $w \geq -1$. 
Thus, below we consider the cases with $w \geq -1$.

\subsection{$-1 \leq w < -1/3$}
\label{subsec:4-1}
\begin{figure*}[htbp]
\begin{minipage}[t]{0.40\hsize}
 \begin{center}
  \includegraphics[width=\hsize]{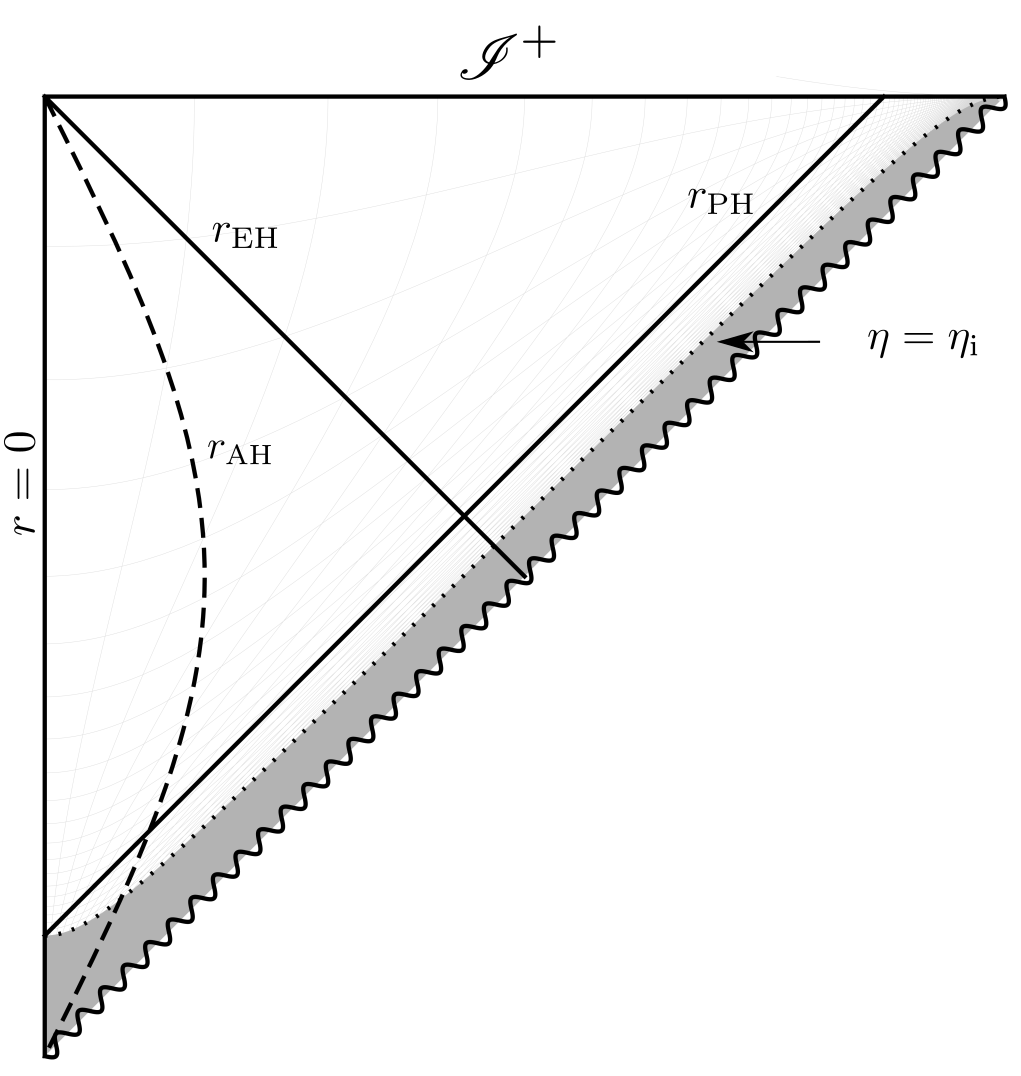}
\caption{Penrose diagram of the accelerated expanding flat FLRW universe with $- 1 < w < - 1/3$:
The gray region is the region excluded by the initial cutoff $\eta_{\text{i}}$. The solid lines express the particle horizon $r_{\text{PH}}$ and the event horizon $r_{\text{EH}}$. 
The dashed curve represents the apparent horizon $r_{\text{AH}}$. 
For $w = -1$ (flat de Sitter universe), the curvature singularity (wavy line) must be replaced by the extendible boundary. 
In this case, the apparent horizon is null and coincides with the event horizon.
}
\label{fig2}
 \end{center}
\end{minipage}
~~
\begin{minipage}[t]{0.57\hsize}
 \begin{center}
  \includegraphics[width=\hsize]{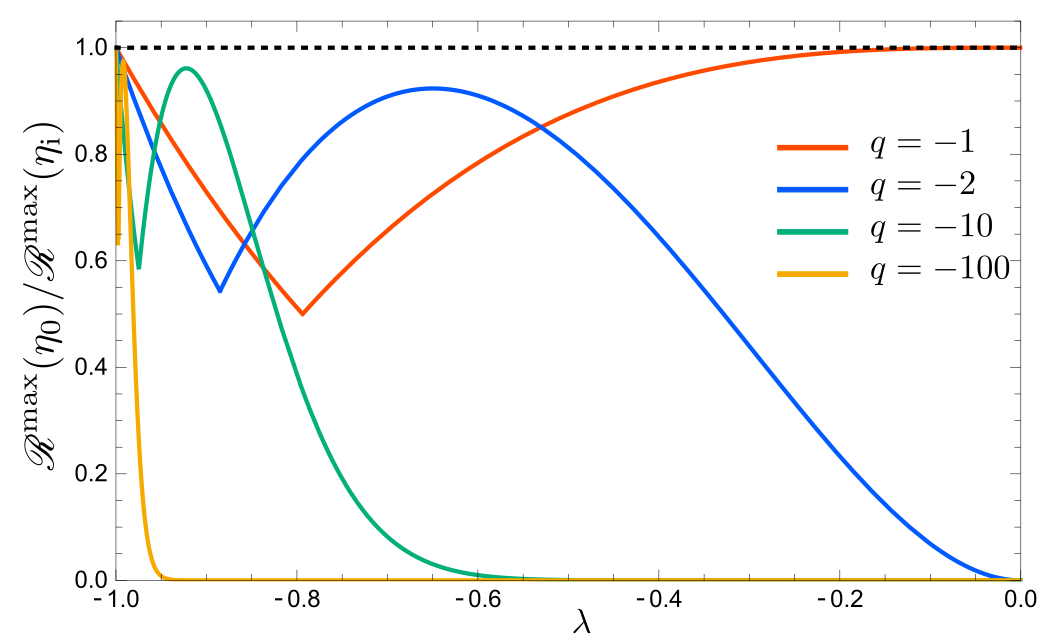}
\caption{Plots of $\mathscr{R}^{\text{max}}(\eta_{0}) / \mathscr{R}^{\text{max}}(\eta_{\text{i}})$ for $ -1 \leq w < -1/3$ $(q \leq -1)$: the plots in different colors represent different values of $q$. 
Every plot is bounded above by $1$ represented by the dashed line. }
\label{fig2_1}
 \end{center}
\end{minipage}
\end{figure*}

Let us consider the case with $-1 \leq w < -1/3$ ($q \leq -1$), which represents the accelerated expanding universe consistent with the null energy condition. 
In this case, the conformal time is defined in $\eta \in (-\infty, 0)$ and the conformal diagram can be written as Fig.~\ref{fig2}.
Since $w < 1$, $\mathscr{R}_{\text{sp}}^{\text{AH}}(\eta_{0})$ diverges at $\eta_{0} \rightarrow - \infty$. Thus, by our criterion, we need to introduce the initial cutoff time $\eta_{\text{i}}$ by $\mathscr{R}_{\text{sp}}^{\text{AH}}(\eta_{\text{i}}) = {\cal O}(1)$. 
We would like to emphasize that this cutoff is introduced independently from the presence of the curvature singularity.
For example, in the case of $w = -1$, $\eta = - \infty$ is a geometrically extendible regular boundary, that is just a coordinate singularity of the flat chart of the de Sitter universe.
Nonetheless, we need a cutoff because, as opposed to the de Sitter universe made by a cosmological constant or a vacuum energy, we are assuming that the universe is made by a thermal fluid with a nonzero entropy current along comoving observers in the flat chart and the entropy current is singular at the geometrically regular boundary.
 
By introducing the initial cutoff $\eta_{\text{i}}$, the reliable region of the spacetime is now defined in $\eta \in (\eta_{\text{i}}, 0)$. 
Note that the particle horizon appears by introducing the initial cutoff, as well as the event horizon:
\begin{align}
 r_{\text{PH}}(\eta) &= \eta - \eta_{\text{i}}, \\
 r_{\text{EH}}(\eta) &= - \eta.
\end{align} 
The apparent horizon is given by
\begin{align}
 r_{\text{AH}}(\eta) &= \frac{\eta}{q},  \qquad (q \leq -1).
\end{align}
The apparent horizon is always smaller than the event horizon, $r_{\text{AH}} < r_{\text{EH}}$, and $r_{\text{PH}}(\eta_{\text{i}})/r_{\text{AH}}(\eta_{\text{i}}) = 0$.
Thus, by Eq.~\eqref{Reta0}, we obtain $\mathscr{R}^{\text{max}}(\eta_{\text{i}}) = \mathscr{R}_{\text{sp}}^{\text{AH}}(\eta_{\text{i}}) = {\cal O}(1)$.

The expression of $\mathscr{R}^{\text{max}}(\eta_{0})$ at an arbitrary time $\eta_0 \in (\eta_{\text{i}},0)$ normalized by the value at $\eta = \eta_{\text{i}}$ can be obtained as 
\begin{widetext}
\begin{align}
\frac{\mathscr{R}^{\text{max}}(\eta_{0})}{\mathscr{R}^{\text{max}}(\eta_{\text{i}})}
&= (-\lambda)^{1 - 2 q} \text{max} 
\left\{
1, q \left( 1 - (-\lambda)^{-1} \right) 
\left[ 3 + 3 q \left( 1 - (-\lambda)^{-1} \right) +  q^2 \left( 1 - (-\lambda)^{-1} \right)^2 \right]
\right\},
\label{eqratio_-1_-1/3}
\end{align}
\end{widetext}
with $\lambda = \eta_{0}/|\eta_{\text{i}}| \in (-1, 0)$. Note that $\lambda$ is increasing from $-1$ to $0$ as $\eta_{0}$ grows. 
The function \eqref{eqratio_-1_-1/3} is plotted in Fig.~\ref{fig2_1}. 
The maximum value of this function is $1$ for any $q \leq -1$.
Hence the entropy-to-area ratio can be evaluated as
\begin{align}
\mathscr{R}^{\text{max}}(\eta_{0}) \leq \mathscr{R}^{\text{max}}(\eta_{\text{i}}) = \mathscr{R}_{\text{sp}}^{\text{AH}}(\eta_{\text{i}}) = {\cal O}(1),
\end{align}
where the last equality follows from the definition of the initial cutoff $\eta_{\text{i}}$ based on our criterion.
This relation tells us that the Bousso bound is satisfied everywhere in the reliable region $\eta_0 \in (\eta_{\text{i}}, 0)$, in the sense $\mathscr{R}^{\text{max}}(\eta_{0}) \leq \mathcal{O}(1)$, once 
 the cutoff $\eta_{\text{i}}$ is introduced by our criterion.

\subsection{$-1/3 < w < 1$}
\begin{figure*}[htbp]
\begin{minipage}[t]{0.4\hsize}
 \begin{center}
  \includegraphics[width= \hsize]{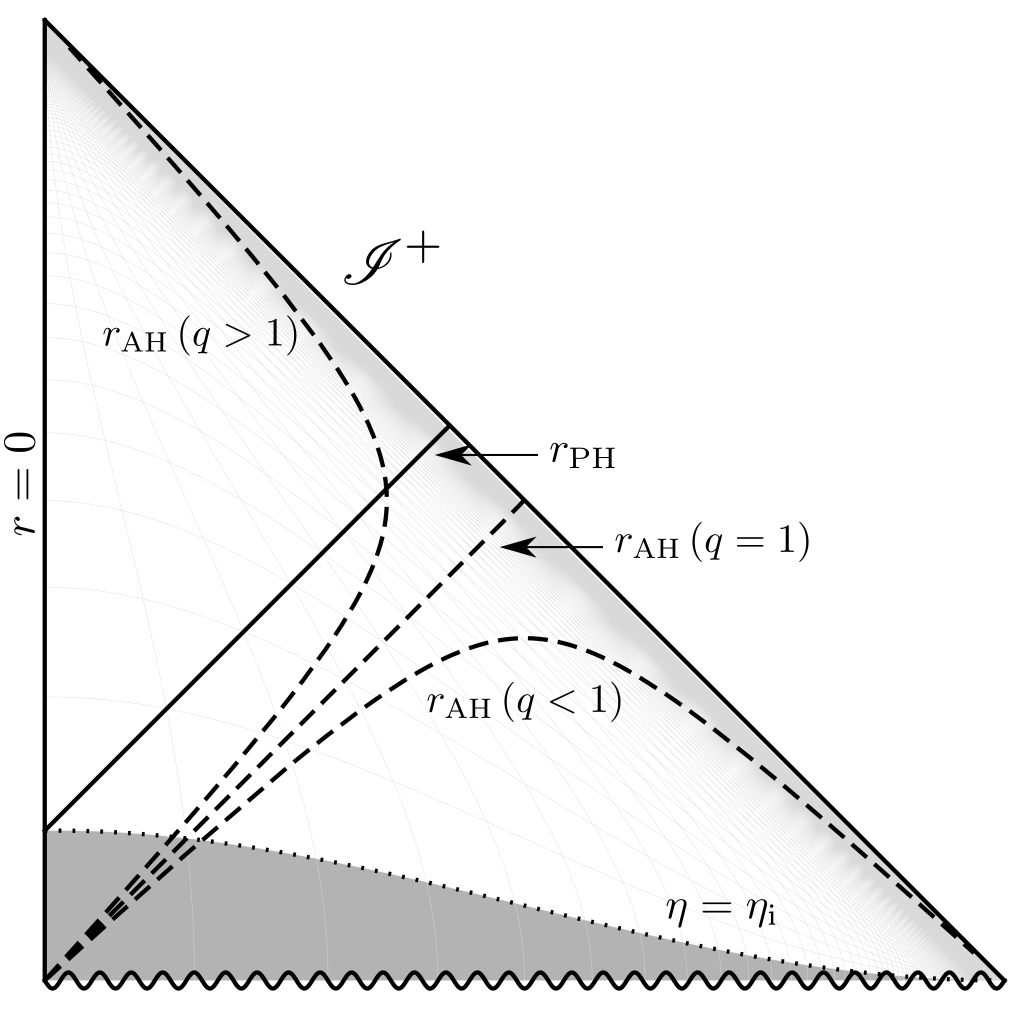}
\caption{Penrose diagram of the decelerated expanding flat FLRW universe with $- 1/3 < w < 1$:
The gray region is the region excluded by the initial cutoff $\eta_{\text{i}}$. The solid line expresses the particle horizon $r_{\text{PH}}$. 
Depending on the value of the parameter $q$, the apparent horizon $r_{\text{AH}}$ becomes timelike, null, or spacelike, which are expressed by the dashed curves.}
\label{fig1}
 \end{center}
\end{minipage}
~~
\begin{minipage}[t]{0.57\hsize}
 \begin{center}
  \includegraphics[width= \hsize]{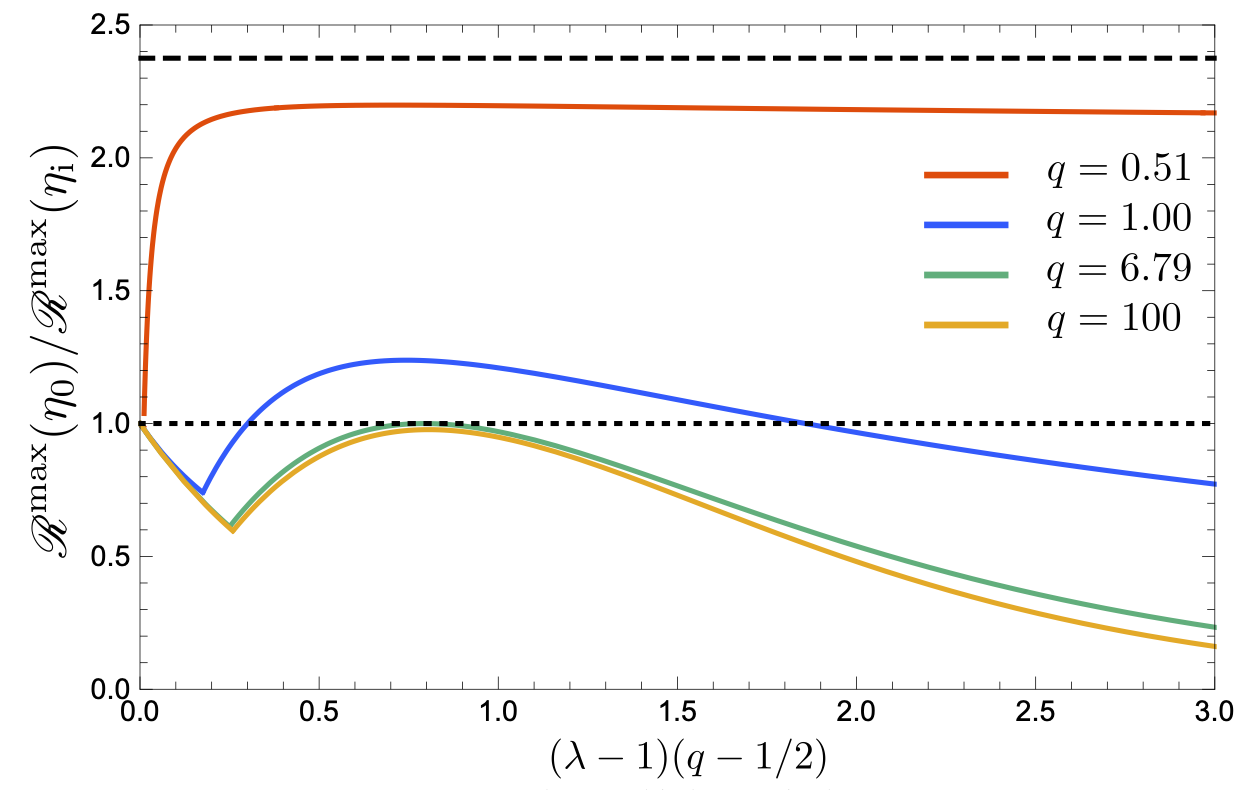}
\caption{Plots of $\mathscr{R}^{\text{max}}(\eta_{0}) / \mathscr{R}^{\text{max}}(\eta_{\text{i}})$ for $-1/3 < w < 1$ $(q > 1/2)$ :the plots in different colors represent different values of $q$. For $q \gtrsim 6.79$, the maximum value is $1$. For $1/2 < q \lesssim 6.79$, the maximum value is greater than $1$ but it is bounded above by $19/8$, which is represented by the bold dashed line. 
}
\label{fig1_1}
 \end{center}
\end{minipage}
\end{figure*}

Next, we consider the decelerated expanding universe fulfilled with a fluid with a constant equation-of-state parameter $- 1/3 < w < 1$, which corresponds to $q > 1/2$. 
This fluid satisfies both the strong energy condition and the dominant energy condition. The conformal time runs $\eta \in (0, \infty)$, and hence the universe is conformally isometric to the upper half of the Minkowski space.
Since the Hubble parameter diverges at $\eta \rightarrow 0$, there corresponds to the initial scalar curvature singularity.
Since $w < 1$, $\mathscr{R}_{\text{sp}}^{\text{AH}}$ also diverges at $\eta \rightarrow 0$.
By our criterion, we need to introduce the initial cutoff time $\eta_{\text{i}}$ by $\mathscr{R}_{\text{sp}}^{\text{AH}}(\eta_{\text{i}}) = {\cal O}(1)$. The Penrose diagram can be drawn as Fig.~\ref{fig1}. 
There is the particle horizon $r_{\text{PH}}(\eta) = \eta - \eta_{\text{i}}$ but there is no event horizon $r_{\text{EH}} = \infty$.

On the cutoff time slice $\eta = \eta_{\text{i}}$, the particle horizon is located at $r_{\text{PH}}(\eta_{\text{i}}) = 0$ but the apparent horizon is located at $r_{\text{AH}}(\eta_{\text{i}}) \neq 0$. 
Hence, the maximum value of the entropy-to-area ratio $\mathscr{R}^{\text{max}}(\eta_{\text{i}})$ reduces to $\mathscr{R}_{\text{sp}}^{\text{AH}}(\eta_{\text{i}})$, which should be ${\cal O}(1)$ by our criterion to introduce the cutoff time.

The maximum value of the entropy-to-area ratio on a time slice $\eta = \eta_0 \in (\eta_{\text{i}}, \infty)$ normalized by the value on the initial time slice $\eta = \eta_{\text{i}}$ can be evaluated as
\begin{widetext}
\begin{align}
\frac{\mathscr{R}^{\text{max}}(\eta_{0})}{\mathscr{R}^{\text{max}}(\eta_{\text{i}})}
= \lambda^{1 - 2 q}
\text{max} \left\{
1, q \left( 1 - \lambda^{-1} \right) 
\left[ 3 + 3 q \left( 1 - \lambda^{-1} \right) +  q^2 \left( 1 - \lambda^{-1} \right)^2 \right]
\right\},
\label{eqratio_-1/3_1}
\end{align}
\end{widetext}
with $\lambda \coloneqq  \eta_{0}/\eta_{\text{i}} \in (1, \infty)$. 
The function \eqref{eqratio_-1/3_1} is plotted in Fig.~\ref{fig1_1}. 
The plots show that $\mathscr{R}^{\text{max}}(\eta_{0})/\mathscr{R}^{\text{max}}(\eta_{\text{i}})$ is bounded above by $19/8$. 
Thus we obtain  
\begin{align}
\mathscr{R}^{\text{max}}(\eta_{0}) < \frac{19}{8} \mathscr{R}^{\text{max}}(\eta_{\text{i}}) = \frac{19}{8} \mathscr{R}_{\text{sp}}^{\text{AH}}(\eta_{\text{i}}) = {\cal O}(1).
\label{bound_-1/3_1}
\end{align}
This relation shows that the universe satisfies the Bousso bound in the sense $\mathscr{R}^{\text{max}}(\eta_{0}) \leq {\cal O}(1)$ for any time $\eta_{0} \geq \eta_{\text{i}}$ once the initial cutoff $\eta_{\text{i}}$ is introduced.

We would like to emphasize that the result \eqref{bound_-1/3_1} is nontrivial. The maximum value of the entropy-to-area ratio for the Bousso bound at any time slice $\eta = \eta_{0}$, $\mathscr{R}^{\text{max}}(\eta_{0})$, is bounded above in terms of the ratio for \textit{the spatial entropy bound at the initial time} $\eta = \eta_{\text{i}}$, $\mathscr{R}_{\text{sp}}^{\text{AH}}(\eta_{\text{i}})$.
Comparing the ratio for the Bousso bound to the ratio for spatial entropy bound on the same time slice $\eta = \eta_{0}$, we obtain (see Eq.~\eqref{Reta0})
\begin{align}
 \mathscr{R}^{\text{max}}(\eta_{0}) \sim q (3 + 3 q + q^2) \mathscr{R}_{\text{sp}}^{\text{AH}}(\eta_{0}).
\end{align} 
Note that the factor $q (3 + 3 q + q^2)$ becomes arbitrarily large when the equation-of-state parameter $w$ is close to $-1/3$. In this case, our criterion applied to the $\eta = \eta_{0}$ slice, that is, the requirement $\mathscr{R}_{\text{sp}}^{\text{AH}}(\eta_{0}) \leq \mathcal{O}(1)$ does \textit{not} guarantee the validity of the Bousso bound. Nonetheless, our analysis shows that the Bousso bound at the $\eta = \eta_{0}$ slice is guaranteed by our requirement at the initial time slice $\eta = \eta_{\text{i}}$.

\subsection{$ w > 1$}
\begin{figure*}[htbp]
\begin{minipage}[t]{0.4\hsize}
 \begin{center}
  \includegraphics[width= \hsize]{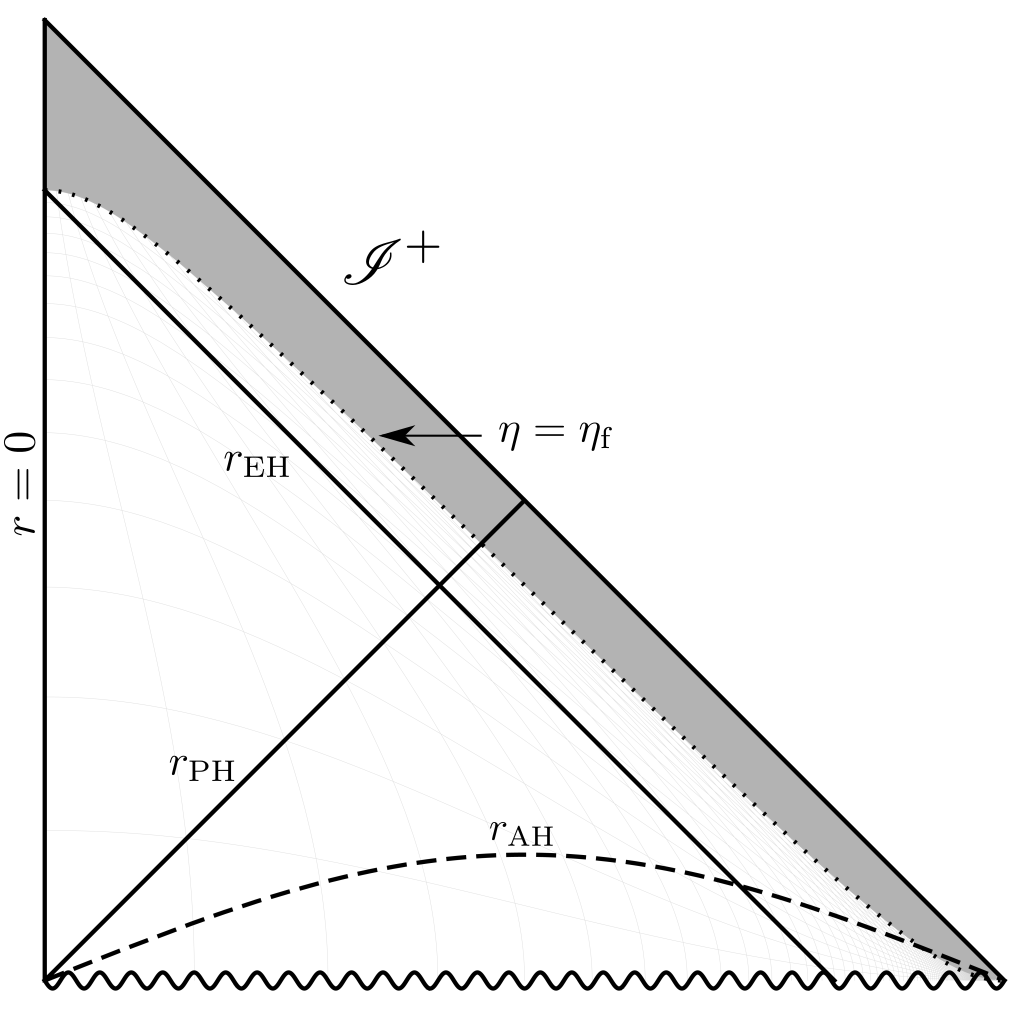}
\caption{Penrose diagram of the decelerated expanding flat FLRW universe with $w > 1$:
the gray region is the region excluded by the final cutoff $\eta_{\text{f}}$.
The solid lines express the particle horizon $r_{\text{PH}}$ and the event horizon $r_{\text{EH}}$. 
The dashed curve represents the apparent horizon $r_{\text{AH}}$. }
\label{fig3}
 \end{center}
\end{minipage}
~~
\begin{minipage}[t]{0.57\hsize}
 \begin{center}
  \includegraphics[width=\hsize]{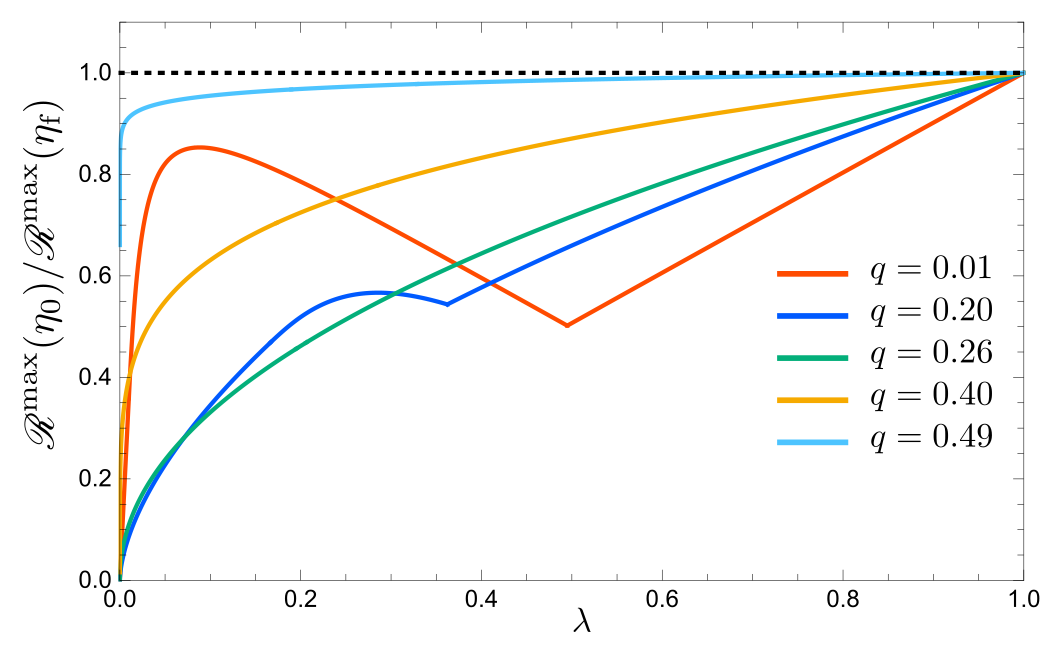}
\caption{Plots of $\mathscr{R}^{\text{max}}(\eta_{0}) / \mathscr{R}^{\text{max}}(\eta_{\text{f}})$ for $w > 1$ $(1 < q < 1/2)$:
the plots in different colors represent different values of $q$. 
Every plot is bounded above by $1$ represented by the dashed line. }
\label{fig3_1}
 \end{center}
\end{minipage}
\end{figure*}

Let us investigate the case of $w > 1$ ($0< q <1/2$), where the dominant energy condition is violated. The conformal time is defined in $\eta \in (0, \infty)$, and the Penrose diagram is given as Fig.~\ref{fig3}. 
Since $\mathscr{R}_{\text{sp}}^{\text{AH}}(\eta_{0})$ diverges at $\eta_{0} \rightarrow \infty$, we need to introduce the future cutoff $\eta_{\text{f}}$ by $\mathscr{R}_{\text{sp}}^{\text{AH}}(\eta_{\text{f}}) = \mathcal{O}(1)$. As a result, we have both the event horizon and the particle horizon:
\begin{align}
 r_{\text{EH}} &= - \eta + \eta_{\text{f}}, \notag\\
 r_{\text{PH}} &= \eta.
\end{align}
We also have the apparent horizon $r_{\text{AH}} = \eta/q$ as usual.

Since there is the initial curvature singularity, it is natural to introduce an initial cutoff $\eta_{\text{i}}$. 
However, to calculate the entropy-to-area ratio, the presence of the curvature singularity itself is not an obstacle to completing the analysis. For this reason, we will assume $\eta_{\text{i}} = 0$ in the following. 

On the future cutoff slice $\eta = \eta_{\text{f}}$, the apparent horizon $r_{\text{AH}}$ and the particle horizon $r_{\text{PH}}$ are finite but the event horizon vanishes by the definition. 
Hence we obtain
\begin{align}
 \mathscr{R}^{\text{max}}(\eta_{\text{f}}) = q ( 3 + 3 q + q^2 ) \mathscr{R}_{\text{sp}}^{\text{AH}}(\eta_{\text{f}}),
\end{align} 
where we used $r_{\text{PH}}/r_{\text{AH}} = q$. 
Since $q \in (0, 1/2)$, the factor is bounded as
\begin{align}
 0< q \left(3 + 3 q + q^2 \right) < \frac{19}{8}.
\end{align}
Thus, it is $\mathcal{O}(1)$.

The expression of $\mathscr{R}^{\text{max}}(\eta_{0})$, i.e., the maximum value of the entropy-to-area ratio on the $\eta = \eta_{0}$ surface, with normalizing the value on $\eta = \eta_{\text{f}}$, can be evaluated as 
\begin{widetext}
\begin{align}
\frac{\mathscr{R}^{\text{max}}(\eta_{0})}{\mathscr{R}^{\text{max}}(\eta_{\text{f}})} =  
\begin{dcases}
 \lambda^{1 - 2 q} \text{max} \left\{
\frac{1}{q ( 3 + 3q + q^2)}, 1  
\right\}, 
&  \left( 0 < \lambda \leq \frac{q}{1+q} \right),  \\
\lambda^{1 - 2 q} \text{max} \left\{
\frac{(-1 + \lambda^{-1})[3 - 3 (-1 + \lambda^{-1})q + (-1 + \lambda^{-1})^2 q^2]}{3 + 3q + q^2}, 1
\right\},
&  \left(\frac{q}{1+q} < \lambda < 1 \right).
\end{dcases}
\label{eqratio_1}
\end{align}
\end{widetext}
Here $\lambda$ is defined by $\lambda \coloneqq  \eta_{0}/\eta_{\text{f}} \in (0, 1)$.
The function \eqref{eqratio_1} is plotted as Fig.~\ref{fig3_1}. One can check that if $q > q_{*} \simeq 0.260$, it can be written by a single function:
\begin{align}
\frac{\mathscr{R}^{\text{max}}(\eta_{0})}{\mathscr{R}^{\text{max}}(\eta_{\text{f}})} = \lambda^{1-2q},
\end{align}
which means that the entropy-to-area ratio of the past-outgoing light sheets is always greater than that of the future-ingoing light sheets. 
In any case, the function \eqref{eqratio_1} is bounded above by $1$, and hence we obtain
\begin{align}
\mathscr{R}^{\text{max}}(\eta_{0}) & \leq \mathscr{R}^{\text{max}}(\eta_{\text{f}}) = q (3 + 3q +q^2) \mathscr{R}_{\text{sp}}^{\text{AH}}(\eta_{\text{f}})
\notag\\
& < \frac{19}{8} \mathscr{R}_{\text{sp}}^{\text{AH}}(\eta_{\text{f}}) = {\cal O}(1),
\end{align}
where the last equality follows from the definition of $\eta_\text{f}$ based on our criterion. 
This relation shows that the Bousso bound is satisfied for any $\eta_0 \in (0, \eta_{\text{f}})$ in the sense $\mathscr{R}^{\text{max}}(\eta_{0}) \leq \mathcal{O}(1)$ once the cutoff $\eta_\text{f}$ is introduced by our criterion.

\subsection{$w = 1$}
Let us consider the case of $w = 1$ $(q = 1/2)$.
In this case, $\mathscr{R}_{\text{sp}}^{\text{AH}}$ is constant in time; see Eq.~\eqref{RspAHs}.
Thus, our criterion leads to the upper bound of the entropy density itself,
\begin{align}
 \mathscr{R}_{\text{sp}}^{\text{AH}} = \frac{1}{2\sqrt{3} \pi} \frac{\bar{s}/M_{\text{pl}}^3}{\sqrt{\bar{\rho}/M_{\text{pl}}^4}} \leq {\cal O}(1),
 \label{RspAH_1}
\end{align}
not the presence of a cutoff time.
The Penrose diagram of this universe is the same as the case $-1/3 < w < 1$ (Fig.~\ref{fig1}) and $w > 1$ (Fig.~\ref{fig3}), except for the absence of the past/future cutoff.
Then the particle horizon is given by
\begin{align}
 r_{\text{PH}}(\eta) = \eta,
\end{align}
and there is no event horizon, $r_{\text{EH}} = \infty$.
The apparent horizon is given by
\begin{align}
 r_{\text{AH}}(\eta) = 2 \eta.
\end{align}
An important point here is that the ratio $r_{\text{PH}}/r_{\text{AH}}$ is constant:
\begin{align}
 \frac{r_{\text{PH}}(\eta)}{r_{\text{AH}}(\eta)} = \frac{1}{2}.
\end{align} 
Since $2^{-1}(3 + 3 \cdot 2^{-1} + 2^{-2}) = 19/8  > 1$, from Eq.~\eqref{Reta0}, we obtain
\begin{align}
 \mathscr{R}^{\text{max}}(\eta_{0}) = \frac{19}{8} \mathscr{R}_{\text{sp}}^{\text{AH}}(\eta_{0}) \leq \mathcal{O}(1),
\end{align}
where the last inequality follows from our criterion \eqref{RspAH_1}.
Hence, the Bousso bound is always satisfied in the sense $\mathscr{R}^{\text{max}}(\eta_{0}) \leq \mathcal{O}(1)$ if our criterion \eqref{RspAH_1} is satisfied.

\subsection{$w = - 1/3$}
\begin{figure*}[htbp]
\begin{minipage}[t]{0.4\hsize}
 \begin{center}
\includegraphics[width= 0.8\hsize]{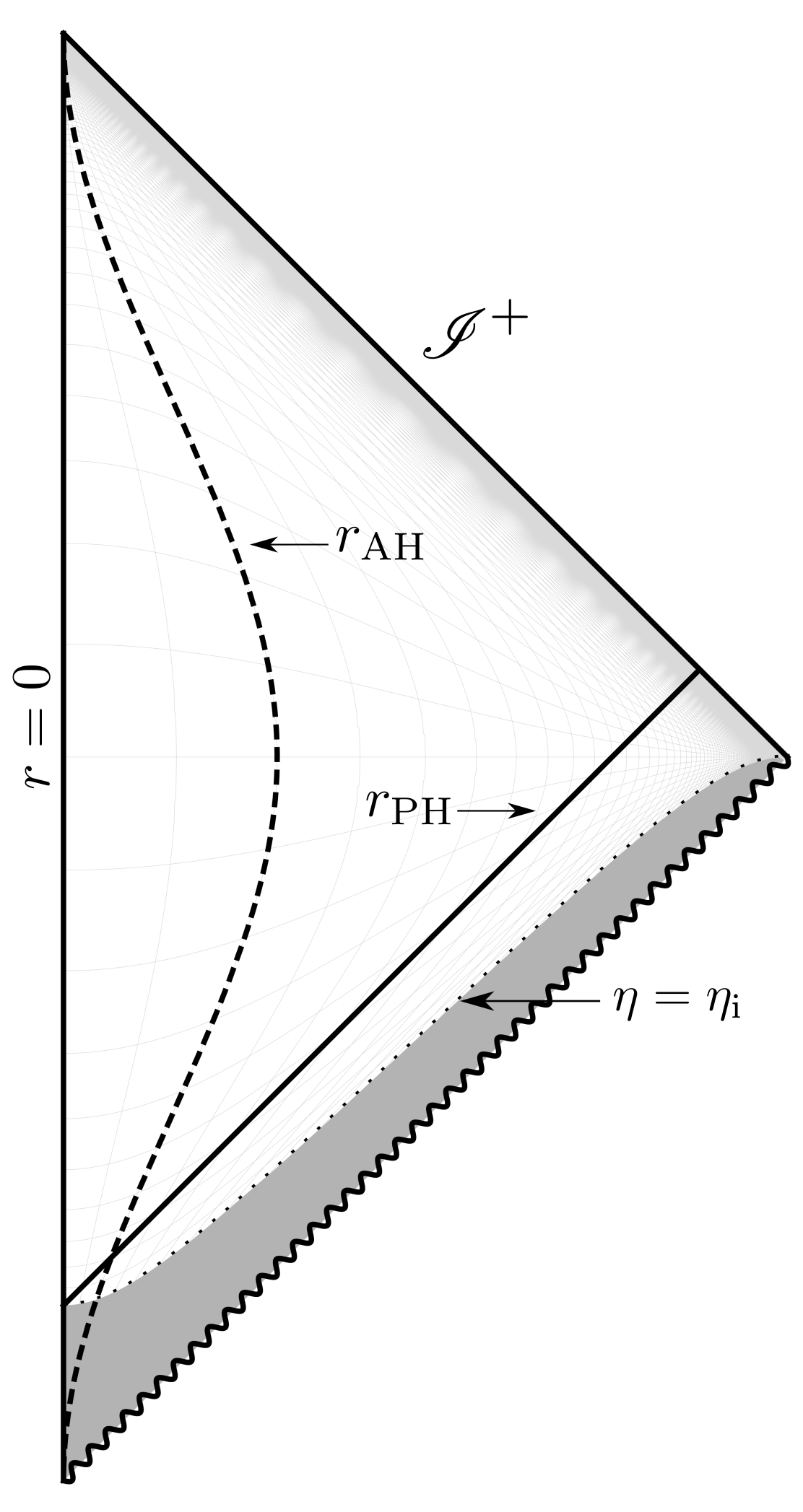}
\caption{Penrose diagram of the expanding flat FLRW universe with a uniform velocity $w  = - 1/ 3$:
the gray region is the region excluded by the initial cutoff $\eta_{\text{i}}$.
The solid line expresses the particle horizon $r_{\text{PH}}$. 
The dashed curve represents the apparent horizon $r_{\text{AH}}$. }
\label{fig5}
 \end{center}
\end{minipage}
~~
\begin{minipage}[t]{0.57\hsize}
 \begin{center}
  \includegraphics[width=\hsize]{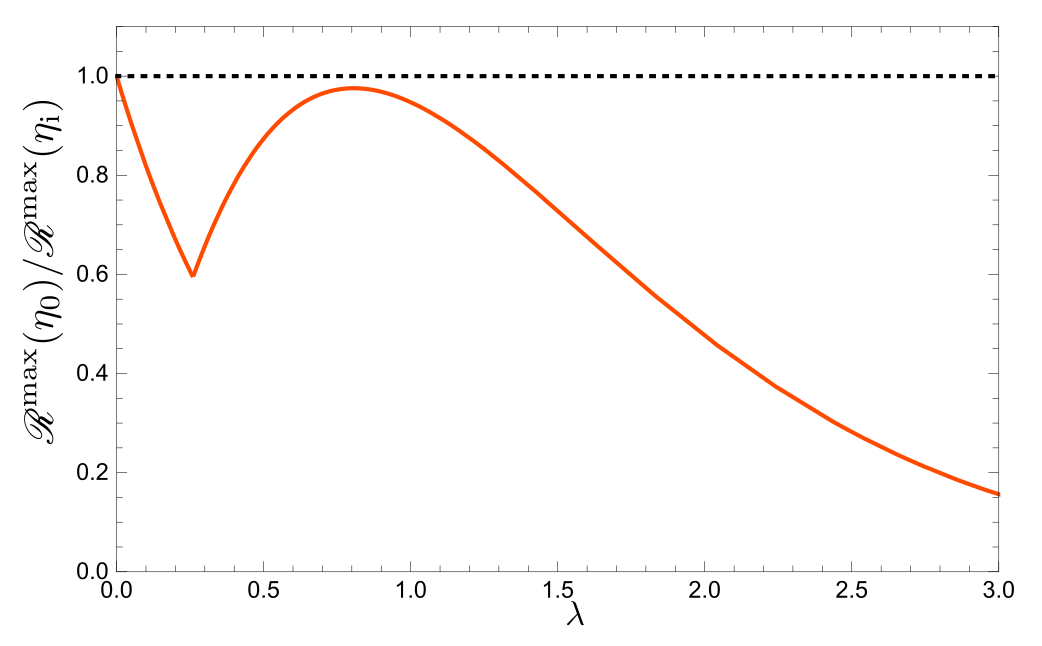}
\caption{
Plot of $\mathscr{R}^{\text{max}}(\eta_{0}) / \mathscr{R}^{\text{max}}(\eta_{\text{i}})$ for $ w = - 1/3$ : the maximum value is bounded above by $1$.
}
\label{fig5_1}
 \end{center}
\end{minipage}
\end{figure*}

Finally, let us consider the case of $w = - 1/3$, where the universe is expanding with a uniform velocity. The energy density and the scale factor can be represented as 
\begin{align}
 \rho = \frac{\bar{\rho}}{a^2}, \qquad a = \mathrm{e}^{\eta/\bar{\eta}},
\end{align}
with a constant $\bar{\eta}$ defined by
\begin{align}
 \bar{\eta} \coloneqq   \sqrt{\frac{3}{8 \pi G \bar{\rho}}}.
\end{align}
Since $\eta$ is defined in $\eta \in (-\infty, \infty)$, the spacetime is conformally isometric to the whole of Minkowski spacetime.
The Penrose diagram of this spacetime is shown in Fig.~\ref{fig5}.
The Hubble parameter can be evaluated as
\begin{align}
 H = \frac{1}{\bar{\eta}}\mathrm{e}^{- \eta/\bar{\eta}}.
\end{align}
Since the Hubble parameter diverges as $\eta \rightarrow - \infty$, there corresponds to a scalar curvature singularity.

Since $\mathscr{R}_{\text{sp}}^{\text{AH}}$ scales as $a^{-2}$ from Eq.~\eqref{RspAHs}, we need to introduce an initial cutoff time $\eta_{\text{i}}$ by $\mathscr{R}_{\text{sp}}^{\text{AH}}(\eta_{\text{i}}) = \mathcal{O}(1)$ based on our criterion.
Then, there is the particle horizon
\begin{align}
 r_{\text{PH}}(\eta) = \eta - \eta_{\text{i}},
\end{align}
whereas there is no event horizon $r_{\text{EH}} = \infty$. The apparent horizon is given by
\begin{align}
 r_{\text{AH}}(\eta) = \bar{\eta}.
\end{align}
Since $r_{\text{PH}}(\eta_{\text{i}})/r_{\text{AH}}(\eta_{\text{i}}) = 0$,
we have $\mathscr{R}^{\text{max}}(\eta_{\text{i}}) = \mathscr{R}_{\text{sp}}^{\text{AH}}(\eta_{\text{i}})$ on the initial time slice.

By introducing $\lambda = (\eta_{0} - \eta_{\text{i}})/\bar{\eta} \in (0, \infty)$, we have
\begin{align}
\frac{\mathscr{R}^{\text{max}}(\eta_{0})}{\mathscr{R}^{\text{max}}(\eta_{\text{i}})} = 
\mathrm{e}^{- 2 \lambda} \text{max} \left\{1 , \lambda (3 + 3\lambda + \lambda^2 ) \right\},
\end{align}
which is plotted in Fig.~\ref{fig5_1}. 
This function has the maximum value $1$ at $\lambda = 0$.
Thus
$\mathscr{R}^{\text{max}}(\eta_{0}) \leq \mathscr{R}^{\text{max}}(\eta_{\text{i}}) = \mathscr{R}_{\text{sp}}^{\text{AH}}(\eta_{\text{i}}) = \mathcal{O}(1)$
and the Bousso bound is satisfied for any $\eta_0 \in (\eta_{\mathrm{i}}, \infty)$ in the sense $\mathscr{R}^{\text{max}}(\eta_{0}) \leq \mathcal{O}(1)$ once the initial cutoff $\eta_{\text{i}}$ is introduced by our criterion.

\section{Nonsingular universe}
\label{sec:5}
In the previous section, we have checked that the Bousso bound is satisfied in any region for any constant equation-of-state parameter that follows the null energy condition $w \geq -1$, once the cutoff time is introduced based on our criterion.
Note that the case $w = -1$ can provide an example of the geometrically nonsingular flat FLRW universe as mentioned in Section \ref{subsec:4-1}.
In this section, we check the validity of the Bousso bound for another nontrivial example of the geometrically nonsingular flat FLRW universe.

To obtain the nonsingular expanding flat FLRW universe, we first need to consider the accelerated expanding universe in the early stage. 
Otherwise, there must be the big bang initial singularity. 
However, by the Borde--Guth--Vilenkin theorem \cite{Borde:2001nh}, there are geodesics incomplete in the past in an accelerated expanding universe. 
One possible way to obtain the nonsingular universe is that the endpoints of such past incomplete geodesics are extendible.
In other words, the initial ``singularity'' of the accelerated expanding flat FLRW universe must be a coordinate singularity like the flat de Sitter spacetime.
The presence or absence of the (scalar and nonscalar) curvature singularity is clarified in Refs.~\cite{Yoshida:2018ndv, Nomura:2021lzz, Nishii:2021ylb}.
There, it is found that the past boundary $a = 0$ is not a curvature singularity when $\partial_{t}H/a^2$ is finite there, though it might be inextendible by a global topological reason \cite{Numasawa:2019juw}. 

\begin{figure*}[htbp]
\begin{minipage}[t]{0.4\hsize}
 \begin{center}
\includegraphics[width= 0.6\hsize]{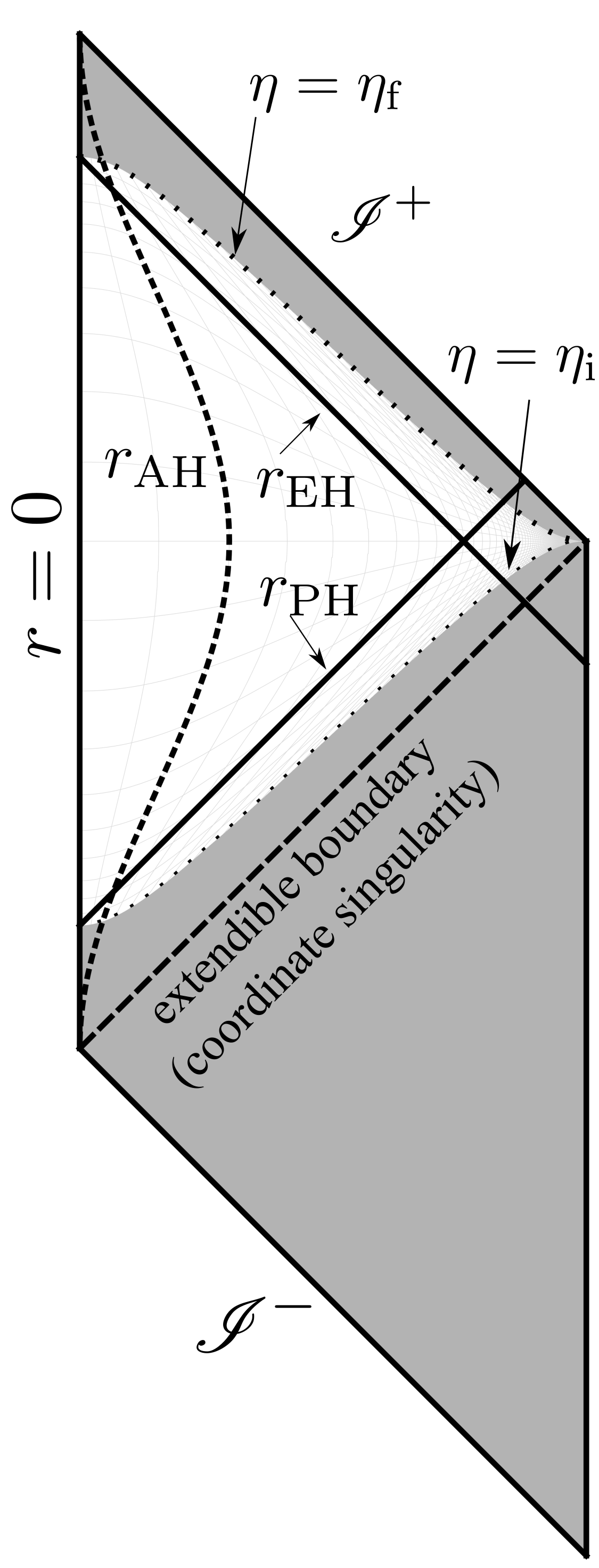}
\caption{Penrose diagram of the geometrically nonsingular universe with the scale factor \eqref{anonsingular}:
The gray region is the region excluded by the initial cutoff $\eta_{\text{i}}$. 
The solid lines express the particle horizon $r_{\text{PH}}$ and the event horizon $r_{\text{EH}}$. The dashed curve represents the apparent horizon $r_{\text{AH}}$.
The long dashed line is the extendible boundary.
There, the scale factor vanishes $a = 0$ but it is the coordinate singularity. The original coordinate system only covers the triangle region above the extendible boundary. Below the boundary, there is a contracting universe that can be obtained by flipping the time from the original metric. 
Now the entire contracting region is excluded by our cutoff.
}
\label{fig6}
 \end{center}
\end{minipage}
~~
\begin{minipage}[t]{0.57\hsize}
 \begin{center}
  \includegraphics[width=\hsize]{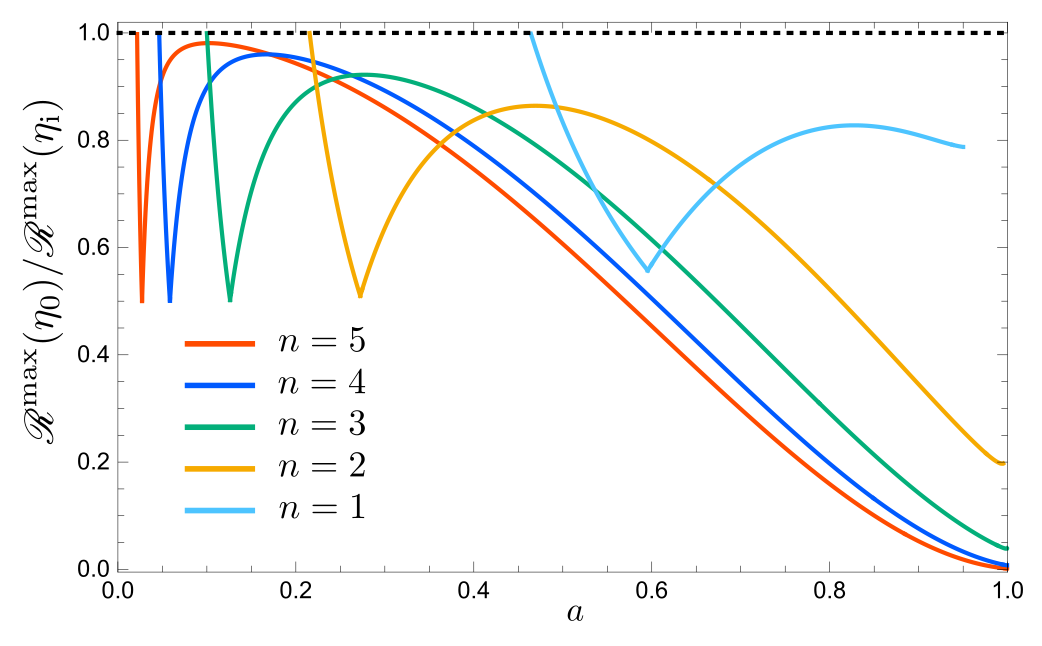}
\caption{Plots of $\mathscr{R}^{\text{max}}(\eta_{0}) / \mathscr{R}^{\text{max}}(\eta_{\text{i}})$ for the geometrically nonsingular universe: the plots in different colors represent different choices of the cutoff. In any case, the maximum value is $1$.}
\label{fig6_1}
 \end{center}
\end{minipage}
\end{figure*}

Here we focus on a simple and analytic example of the geometrically nonsingular flat FLRW universe following the null energy condition investigated in Ref.~\cite{Nomura:2022vcj}, that has topologically $S^{3}$ Cauchy surfaces after the maximal extension beyond the initial coordinate singularity at $a =0$.
The scale factor of this universe, defined only in the flat chart, is given by
\begin{align}
 a(t) = \frac{\mathrm{e}^{\bar{H} t}}{\sqrt{1 + \mathrm{e}^{2 \bar{H} t}}}, \label{anonsingular}
\end{align}
where $\bar{H}$ is a positive constant.
The Penrose diagram of this universe is shown in Fig.~\ref{fig6}.

The first remark is that, based on our criterion \eqref{ourcriterion}, we need to introduce the past cutoff on the flat chart because $\mathscr{R}_{\text{sp}}^{\text{AH}}$ diverges at $a \rightarrow 0$.
This is because $a = 0$ is geometrically nonsingular and hence $\rho$ is finite there, but the adiabatic entropy current diverges there, $s \rightarrow \infty$.
This is a common property of geometrically nonsingular flat FLRW universes that follow the null energy condition.
Note that, regarding the spacetime region after introducing the initial cutoff on the flat chart as the whole spacetime, it has Cauchy surfaces with the topology $R^3$, not $S^3$.

In addition, with our specific choice of the scale factor \eqref{anonsingular}, the dominant energy condition is violated in a sufficient future. 
Actually, $\mathscr{R}_{\text{sp}}^{\text{AH}}$ diverges in the infinite future and we also need to introduce the future cutoff $\eta_{\text{f}}$.

Let us calculate the maximum value of the entropy-to-area ratio $\mathscr{R}^{\text{max}}$.
Since our scale factor $a \in (0,1)$ is a monotonically increasing function in time, let us use $a$ as the time. 
The Hubble parameter can be expressed as 
\begin{align}
 H(a) = ( 1 -  a^2) \bar{H}.
\end{align}
Then, we get the expression of $\mathscr{R}_{\text{sp}}^{\text{AH}}$ as a function of $a$,
\begin{align}
 \mathscr{R}_{\text{sp}}^{\text{AH}}(a) = \frac{\bar{s}}{6 \pi \bar{H} M_{\text{pl}}^2} \frac{1}{a^3(1 - a^2)}.
\end{align}
From our criterion, the cutoff times are introduced by $\mathscr{R}_{\text{sp}}^{\text{AH}}(a_{\text{i}}) = {\cal O}(1)$ and $\mathscr{R}_{\text{sp}}^{\text{AH}}(a_{\text{f}}) = {\cal O}(1)$.
Assuming $a_{\text{i}} \ll 1$ and $1 - a_{\text{f}} \ll 1$, we obtain
\begin{align}
 a_{\text{i}} &\sim {\cal O}(1) \cdot \left(\frac{\bar{s}}{6 \pi \bar{H}M_{\text{pl}}^2}\right)^{1/3},\label{ai}\\
 a_{\text{f}}&\sim 1 - {\cal O}(1) \cdot \frac{1}{2} \frac{\bar{s}}{6 \pi \bar{H}M_{\text{pl}}^2} .
 \label{af}
\end{align}

From the expression of $H$, we can write the apparent horizon as
\begin{align}
 r_{\text{AH}}(a) &= \frac{1}{a(1 - a^2) \bar{H}}.
\end{align}
By the definition of the conformal time, we can express the conformal time as a function of $a$ as
\begin{align}
 \eta(a) = - \frac{1}{a \bar{H}} + \frac{1}{\bar{H}} \text{Arctanh}(a).
\end{align}
Then, the event horizon and the particle horizon can be expressed as functions of $a$,
\begin{align}
 r_{\text{EH}}(a) &= \frac{1}{\bar{H}} \left(\frac{1}{a} - \frac{1}{a_{\text{f}}} - \text{Arctanh}(a) + \text{Arctanh}(a_{\text{f}})\right), \\
r_{\text{PH}}(a) &= - \frac{1}{\bar{H}} \left(\frac{1}{a} - \frac{1}{a_{\text{i}}} - \text{Arctanh}(a) + \text{Arctanh}(a_{\text{i}})\right).
\end{align}
Since $r_{\text{PH}}(a_{\text{i}}) =0$ and $r_{\text{AH}}(a_{\text{i}}) \neq 0$, we obtain $\mathscr{R}^{\text{max}}(a_{\text{i}}) = \mathscr{R}^{\text{AH}}_{\text{sp}}(a_{\text{i}})$, and hence the Bousso bound is satisfied at the initial cutoff slice.

Then the plots of $\mathscr{R}^{\text{max}}(a)/ \mathscr{R}^{\text{max}}(a_{\text{i}})$ are given in Fig.~\ref{fig6_1}.
Here the cutoffs are chosen as
\begin{align}
a_{\text{i}} =  10^{- n/3}, \qquad a_{\text{f}} = 1 - \frac{1}{2} \times 10^{- n}, 
\label{nonsing_cut}
\end{align}
which corresponds to the choice 
\begin{align}
 \frac{\bar{s}}{6 \pi \bar{H}M_{\text{pl}}^2} = 10^{-n},
\end{align}
and the ${\cal O}(1)$ coefficients in Eqs.~\eqref{ai} and \eqref{af} are set to unity.
The plots show that $\mathscr{R}^{\text{max}}(a) \leq \mathscr{R}^{\text{max}}(a_{\text{i}})$ for any $a \in (a_{\text{i}}, a_{\text{f}})$ and any choice of the cutoff parameter $n$ in Eq.~\eqref{nonsing_cut}.
Hence, the Bousso bound is always satisfied in $a \in (a_{\text{i}}, a_{\text{f}})$ in the sense $\mathscr{R}^{\text{max}}(a) \leq \mathcal{O}(1)$  once the initial cutoff $a_\text{i}$ is introduced based on our criterion $\mathscr{R}_{\text{sp}}^{\text{AH}}(a_\text{i}) = \mathcal{O}(1)$.

\section{Summary and Discussion}
In this paper, we proposed a criterion for introducing a cutoff time for the flat homogeneous and isotropic universe so that Bousso's entropy bound is satisfied over the entire resultant spacetime region. 
Our criterion is expressed by Eq.~\eqref{ourcriterion}, $\mathscr{R}_{\text{sp}}^{\text{AH}}(\eta_{\text{cut}}) = {\cal O}(1)$.
This implies that the spatial entropy bound with respect to the apparent horizon must be satisfied, $\mathscr{R}_{\text{sp}}^{\text{AH}}(\eta) \leq {\cal O}(1)$, for a physically reasonable spacetime region. 
We have explicitly checked that, after introducing the cutoff based on our criterion, the Bousso bound is always satisfied in the universe with a fluid with a constant equation-of-state parameter consistent with the null energy condition. 
In addition, we have investigated the validity of the Bousso bound in an example of the geometrically nonsingular universe and found that our criterion works well even in this case. 
While the criterion used in the previous works \cite{Bousso:1999xy, Kaloper:1999tt} is applicable only near a curvature singularity, our criterion works even for the geometrically nonsingular universe. 
Our result suggests that the incompleteness predicted by the singularity theorem based on the entropy bounds means the presence of a geometrical singularity or a too large amount of entropy as our criterion \eqref{ourcriterion}, at least for the examples of the flat FLRW universe investigated in this paper.
This result is reasonable because the geometrically nonsingular, spatially flat, homogeneous, and isotropic universe that follows the null energy condition is possible only when $a = 0$ is the coordinate singularity \cite{Yoshida:2018ndv}. 
However, with the assumption of the nonzero adiabatic entropy current, $a = 0$ becomes a kind of singularity of the entropy density $s(t) \rightarrow \infty$, even though it is geometrically regular (coordinate singularity). 

Though the spatial entropy bound and the Bousso bound are nonlocal properties of the system, our criterion itself is expressed by local quantities: entropy density and the energy density, thanks to the homogeneity and isotropy of the spacetime. 
Thus it might be able to understand the necessity of the cutoff from the viewpoint of local physics, like a cutoff based on a curvature singularity.
If one applies the Bekenstein bound \eqref{Bekensteinbound} for a Planck volume, $S \sim s M_{\text{pl}}^{-3}$, $E \sim \rho M_{\text{pl}}^{-3}$ and $R \sim M^{-1}_{\text{pl}}$, we obtain $ s M_{\text{pl}}^{-3} < \# \cdot \rho M_{\text{pl}}^{-4} $ with a numerical factor $\#$. In addition, assuming the energy density is below the Planck scale cutoff, $\rho M_{\text{pl}}^{-4} < 1$, our criterion is automatically satisfied:
\begin{align}
 (s M_{\text{pl}}^{-3})^2 < \# \cdot (\rho M_{\text{pl}}^{-4})^2 < \# \cdot \rho M_{\text{pl}}^{-4}.
\end{align}
Thus the Bekenstein bound for a Planck volume below the Planck energy scale is a sufficient condition for the Bousso bound. 

In this paper, we just checked our criterion with a few examples of the expanding universe consistent with the null energy condition. 
It is interesting to check whether our criterion works well for more general universes and give proof of the Bousso bound from our criterion. 
One possible hint might be the proof of the Bousso bound based on the entropy current given in Ref.~\cite{Flanagan:1999jp}. 
One of the sets of assumptions are
\begin{align}
 (s_{\mu} k^{\mu})^2 &\leq \alpha_{1} T_{\mu\nu} k^{\mu} k^{\nu}, \\
|k^{\mu} k^{\nu} \nabla_{\mu} s_{\nu}| &\leq \alpha_{2} T_{\mu\nu} k^{\mu} k^{\nu},
\end{align}
for any null vector $k^{\mu} \partial_{\mu}$, where the positive constants $\alpha_{1}$ and $\alpha_{2}$ are assumed to satisfy  $(\pi \alpha_{1})^{1/4} + (\alpha_{2}/\pi)^{1/2} = 1$.
If the null vector $k^{\mu} \partial_{\mu}$ is replaced with a timelike vector, the above requirement would reduce to the inequality that states the square of the entropy density should be smaller than energy. That is basically nothing but our criterion. 

\begin{acknowledgments}
K.N.~was supported by Grant-in-Aid for JSPS Research Fellowship and JSPS KAKENHI Grant No.JP21J20600.
D.Y.~was supported by JSPS KAKENHI Grants No.JP20K14469 and No.JP21H05189.
\end{acknowledgments}

\appendix

\bibliography{ref}
\end{document}